\documentclass[12pt]{article}
\usepackage{a4}
\usepackage{longtable}
\usepackage{epsfig}
\usepackage{cite}
\newcommand{\be}{\begin{equation}}
\newcommand{\ee}{\end{equation}}

\newcommand{\Gb}{\overline{G}}

\newcommand{\phicl}{\phi_{\mbox{\tiny cl}}}

\newcommand{\nn}{\nonumber \\}

\newcommand{\beq}{\begin{equation}}
\newcommand{\eeq}{\end{equation}}
\newcommand{\bea}{\begin{eqnarray}}
\newcommand{\eea}{\end{eqnarray}}
\newcommand{\beqan}{\begin{eqnarray*}}
\newcommand{\eeqan}{\end{eqnarray*}}
\newcommand{\ba}{\begin{array}}
\newcommand{\ea}{\end{array}}

\newcommand{\bdm}{\begin{displaymath}}
\newcommand{\edm}{\end{displaymath}}
\newcommand{\lgl}{\langle}
\newcommand{\rgl}{\rangle}

\newcommand{\dg}{\dagger}

\newcommand{\ccdot}{\hskip-0.3ex\cdot\hskip-0.3ex}

\begin{document}

\thispagestyle{empty}
\begin{flushright}
LU TP 99-14\\
ZU--TH 18/99\\
UWThPh-1999-42\\
July 1999
\end{flushright}
\vspace{2cm}
\begin{center}
\begin{Large}
{\bf Renormalization of Chiral Perturbation Theory\\[5pt] 
to Order $p^6$*} \\[1cm]
\end{Large}
{\bf J. Bijnens$^1$, G. Colangelo$^2$ and G. Ecker$^3$} \\[2cm]
\end{center}
\begin{flushleft} 
${}^1$ Dept. of Theor. Phys., Univ. Lund, S\"olvegatan 14A, 
S--22362 Lund, Sweden\\
${}^2$ Inst. Theor. Physik, Univ. Z\"urich, Winterthurerstr. 190,
CH--8057 Z\"urich--\\~\, Irchel, Switzerland.\\ 
${}^3$ Inst. Theor. Phys., Univ. Wien, Boltzmanng. 5, A--1090 Wien,
Austria\\[1cm]
\end{flushleft}

\begin{abstract}
\noindent
The renormalization of chiral perturbation theory is carried out to 
next-to-next-to-leading order in the meson sector. We calculate the 
divergent part of the generating functional of Green functions of
quark currents to $O(p^6)$ for chiral $SU(n)$, involving one- and
two-loop diagrams. The renormalization group equations for the
renormalized low-energy constants of $O(p^6)$ are derived. We compare
our results with previous two-loop calculations in chiral perturbation
theory.
\end{abstract}
\setcounter{page}{0}

\vfill
\noindent * Work supported in part by TMR, EC-Contract No. 
ERBFMRX-CT980169 \\(EURODA$\Phi$NE).\\

\clearpage

\renewcommand{\theequation}{\arabic{section}.\arabic{equation}}
\section{Introduction}
Chiral perturbation theory (CHPT) \cite{Wein79,GL84,GL85} provides a
systematic low-energy expansion of QCD. In the meson sector, this
expansion is now being carried out to next-to-next-to-leading order 
\cite{badhonnef}. 

In this paper we perform the complete renormalization of the
generating functional of Green functions of quark currents to
$O(p^6)$. As in every quantum field theory with a local action, the
divergences of $O(p^6)$, due to one- and two-loop diagrams, are
themselves local so that the functional can be rendered finite by 
the chiral action of $O(p^6)$. We carry out the calculation for 
chiral $SU(n)$ and then specialize to the realistic cases $n=2,3$.
The divergent part of the generating functional serves as a check 
for existing and all future two-loop calculations in the meson
sector. Moreover, the divergences govern 
the renormalization group equations for the renormalized low-energy 
constants of $O(p^6)$. The double-pole divergences (in dimensional
regularization) determine the double chiral logs, the leading
infrared singularities of $O(p^6)$ \cite{Wein79,Col95,BCE1}.

The main tools for this calculation are well known. One starts by
expanding the chiral action around the solution of the lowest-order 
equation of motion (EOM) to generate the loop expansion upon
functional integration. We then employ heat-kernel techniques for
extracting the divergences of the generating functional. We 
discuss first a generic quantum field theory of scalar fields before 
specializing to CHPT. Special emphasis is given to certain aspects 
specific to  nonrenormalizable quantum field theories. One subtlety
concerns the dependence of the renormalization procedure on the choice 
of the Lagrangian of next-to-leading order, $O(p^4)$. The allowed
Lagrangians differ by EOM terms. As a consequence, the sum of 
one-particle-reducible (1PR) diagrams is in general 
divergent. It is shown that these divergences are always local 
and can therefore also be renormalized by the chiral action of
$O(p^6)$. Likewise, the proper renormalization at $O(p^4)$ guarantees
the absence of nonlocal subdivergences even in the presence of such 
EOM terms.

The major burden of this calculation is the algebraic complexity of
CHPT at $O(p^6)$. We exemplify the laborious calculations by three
special cases of one-particle irreducible (1PI) diagrams. The results 
for $SU(n)$ and for $n=2,3$ are collected in several tables with the 
double-and single-pole divergences, using the standard bases of
Ref.~\cite{BCE2} for the chiral Lagrangians of $O(p^6)$. 

The paper is organized as follows. In Sec.~\ref{sec:loops}, we
describe the various steps of the calculation for a general scalar
quantum field theory. The divergences of 1PR
diagrams associated with EOM terms in the Lagrangian of next-to-leading
order are shown to be local. Nonlocal subdivergences cancel, leading in
particular to a set of consistency conditions for the double-pole
coefficients of the 1PI diagrams \cite{Wein79}. 
In the following section, we specialize 
to CHPT and expand the chiral actions of $O(p^2)$ and $O(p^4)$ to the
required orders in the fluctuation fields. Using the short-distance
behaviour of (products of) Green functions \cite{JO}, the
determination of the divergence functional reduces to the calculation
of the appropriate combinations of Seeley-DeWitt coefficients. We
discuss one example each for the one-loop diagrams with a single
vertex of $O(p^4)$ and for the genuine two-loop diagrams (butterfly
and sunset topologies). The renormalization procedure is described in
Sec.~\ref{sec:results} including a derivation of renormalization group
equations for the renormalized low-energy constants of $O(p^6)$. In
Sec.~\ref{sec:app}, we derive the tree-level contributions of the 
$O(p^6)$ Lagrangian introduced in Ref.~\cite{BCE2}
for several processes that have already been calculated to $O(p^6)$.
This allows for an immediate check of the
divergence structure. Sec.~\ref{sec:conc} contains our conclusions.
The singular parts of the products of operators \cite{JO} are
reproduced in App.~\ref{app:JO} and the manipulations for verifying
the cancellation of nonlocal subdivergences are described in 
App.~\ref{app:subdivs}. Our actual results in the form of double- and
single-pole divergences are collected in several tables in
App.~\ref{app:n23} for $n=3$ and $2$ light flavours and in 
App.~\ref{app:su(n)} for chiral $SU(n)$.
For completeness we also give the explicit form of the terms of the
$O(p^6)$ Lagrangian in Apps.~\ref{app:n23} and \ref{app:su(n)}.

\addtocounter{section}{0}
\setcounter{equation}{0}
\section{Loop expansion in quantum field theory}
\label{sec:loops}

In this section we set the general framework for our calculation. Although
the treatment will be valid for any quantum field theory of scalar fields
the notation is tailored to CHPT.  We shall work in Euclidean space both in
the present section and in the following where we describe the actual
calculations. The final result in the form of the divergence functional of
$O(p^6)$ will however be given in Minkowski space in
Sec.~\ref{sec:results}.

\subsection{Generating functional}

The starting point for a quantum field theory is the classical action
$S_2$. We supplement this action by quantum corrections $S_{2n}$ ($n >
1$) and define the total action $S$ as
\be
S=S_2+\hbar S_4 + \hbar^2 S_6 + \ldots = \sum_{n=0}^{\infty} \hbar^n
S_{2+2n} \; \; . 
\ee
The numbering of terms is chosen to comply with CHPT.
Each term depends on a set of $N$ scalar fields $\phi= 
\{\phi_1,\ldots,\phi_N \}$ and on $M$ external sources 
${\bf j}= \{j_1,\ldots,j_M \}$:
\be
S_i=S_i[\phi ,{\bf j}] \; \; .
\ee
Unlike in a renormalizable quantum field theory, the quantum 
corrections $S_{2n}$ ($n > 1$) are in general not of the form of
$S_2$. In fact, in CHPT the $S_{2n}$ do not have any terms in common 
for different $n$ because $S_{2n}$ is of chiral order $p^{2n}$.

The generating functional $Z[\bf j]$ is the logarithm of the 
vacuum-to-vacuum transition amplitude in the presence of the external 
sources. It can be defined via a path integral
\be
\exp(-Z[{\bf j}]/\hbar) = {\cal N}^{-1} \int \prod [d \phi_i] 
e^{- S[\phi,{\bf j}]/\hbar}
\label{pint}
\ee
with the normalization
\be
Z[0]=0 \; \; .
\ee

The loop expansion can be constructed as an
expansion around the solution of the classical EOM:
\be
\frac{\delta S_2}{\delta \phi_i}=0 \; \; \; \Rightarrow \; \; \; 
\phicl [{\bf j}] \; \; .\label{EOM}
\ee 
The action $S[\phi,{\bf j}]$ is now expanded around the classical
field $\phicl [{\bf j}]$,
\be
S[\phi,{\bf j}] = S_2[ \phicl, {\bf j}]+\left( {1 \over 2}S_2^{(2)}[
  \phicl, {\bf j}] \xi^2 + \hbar S_4[ \phicl, {\bf j}] \right)
+S_{\mbox{\tiny int}}[\phi,{\bf j}] \; \; , 
\label{S_h_exp}
\ee 
where $\xi_i = \phi_i-\phi_{\mbox{\tiny cl} \; i}$.
Here and in the sequel we adopt the following compact notation:
\be
S_n^{(m)}=S_{n \,i_1, \ldots, i_m}^{(m)}(x_1,\ldots,x_m) = \frac{\delta^m
  S_n[\phi,{\bf j}]}{\delta \phi_{i_1}(x_1) \ldots \delta
  \phi_{i_m}(x_m)} 
\ee
\be
S_n^{(m)}\xi^m= \int dx_1 \ldots dx_m S_{n \,i_1, \ldots,
  i_m}^{(m)}(x_1,\ldots,x_m) \xi_{i_1}(x_1) 
\ldots \xi_{i_m}(x_m) \; \; .
\ee

The term $S_2[ \phicl, {\bf j}]$ is  the only one that survives
in the limit $\hbar \rightarrow 0$. In Eq.~(\ref{S_h_exp}) we have 
written between parentheses the terms that after functional
integration contribute to $O(\hbar)$. The rest is denoted  
$S_{\mbox{\tiny int}}[\phi,{\bf j}]$. Restricted to terms contributing
to $O(\hbar^2)$, it takes the explicit form
\bea
S_{\mbox{\tiny int}}[\phi,{\bf j}] &=& {1 \over 3!} S_2^{(3)}[ \phicl,
{\bf j}] \xi^3 + {1 \over 4!} S_2^{(4)}[ \phicl, {\bf j}] \xi^4 + \ldots
\nonumber \\ 
&+& \hbar \left( S_4^{(1)}[ \phicl, {\bf j}] \xi +{1 \over 2}S_4^{(2)}[
  \phicl, {\bf j}] \xi^2 + \ldots \right) \nonumber \\
&+& \hbar^2 S_6[ \phicl, {\bf j}] \; \; .
\label{Sint_h_exp}
\eea

Functional integration over the fluctuation fields $\xi_i$ in 
(\ref{pint}) gives rise to an expansion of the generating functional 
$Z[\bf j]$ in powers of $\hbar$,
\be
Z=Z_0+\hbar Z_1 +\hbar^2 Z_2 +O(\hbar^3) \; \; ,
\ee
where
\bea
Z_0 &=& \overline{S_2} \nonumber \; \; , \\
Z_1 &=& \overline{S_4} + \frac{1}{2} \mbox{Tr} \ln \left( {\Delta \over
    \Delta_0 } \right) \nonumber \; \; \\
Z_2 &=& \overline{S_6} +z_2 \; \; .
\label{Z_h_exp}
\eea
$\overline{S_n}=S_n[\phicl,{\bf j}]$ is the action evaluated at the
classical field and $\Delta = S_2^{(2)}[\phicl,{\bf j}]$ is a
differential operator. The subscript $0$ in $\Delta_0$ indicates that
the operator is to be evaluated for ${\bf j}=0$.

The explicit expression for $z_2$ is
\bea
z_2 &=&{1 \over 8} S_{2\; ijkl}^{(4)} G_{ij} G_{kl} - {1 \over 12} S_{2\;
  ijk}^{(3)} G_{ir} G_{js} G_{kt} S_{2\; rst}^{(3)} 
  - {1 \over 8}  S_{2\; ijk}^{(3)} G_{ij} G_{kt} G_{rs} S_{2\;
  rst}^{(3)}\nn
&+& {1 \over 2} S_{4\;ij}^{(2)} G_{ij} 
- {1 \over 2} S_{2\;ijk}^{(3)} G_{jk} G_{il} S_{4\; l}^{(1)}\nonumber \\
&-& {1 \over 2} S_{4\; i}^{(1)} G_{ij} S_{4\; j}^{(1)} \;\; ,
\label{z2}
\eea
where $G_{ij}$ is the inverse of the differential operator $\Delta$:
\be
\Delta_{ij}(x) G_{jk}(x,y) = \delta_{ik} \delta^{(4)}(x-y) \; \; .
\ee
Summation over repeated indices is understood.
In Eq.~(\ref{z2}) we have adopted a compact notation where the Latin
indices stand both for flavour (discrete) and space-time (continuous)
indices. Repeated Latin indices indicate both a summation 
over flavour indices and integration over space-time variables.

\begin{figure}
\centerline{\epsfig{file=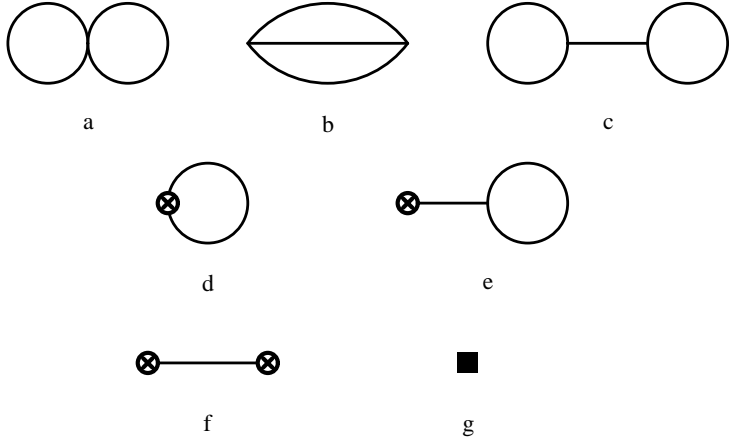,height=6cm}}
\caption{Diagrams contributing to the generating functional $Z_2$
in (\protect\ref{Z_h_exp}).
The propagators and vertices carry the full tree structure associated
with the lowest-order action $S_2$. Normal vertices are from
$S_2$, crossed circles denote vertices from $S_4$. The diagrams are
shown in the order appearing in (\protect\ref{z2}). Diagram
g denotes tree diagrams corresponding to $\overline{S_6}$ in
(\protect\ref{Z_h_exp}).}
\label{fig:p6diag}
\end{figure}

The terms in Eq.~(\ref{z2}) include two-loop (first line), one-loop
(second line) and tree-level contributions (third line, together with
$\overline{S_6}$ in (\ref{Z_h_exp})). In the order written, they
correspond to the diagrams shown in Fig.~\ref{fig:p6diag}.

\subsection{Heat-kernel expansion}

To calculate the divergent part of the generating functional we use the
well-known heat-kernel technique. We sketch here the procedure and
introduce the relevant notation, following closely the presentation of
Ref.~\cite{JO}.

Given an elliptic differential operator $\Delta$,
\be
\Delta(x) = -d_x^2+\sigma(x)\; , \; \; \; d_\mu=\partial_\mu+\gamma_\mu(x)
\; \; , 
\ee
where both $\sigma$ and $\gamma_\mu$ are matrices in general, we define its
heat kernel as the function that satisfies the heat diffusion equation
\be
\left({\partial \over \partial t} + \Delta(x) \right){\cal G}_\Delta
(x,y;t) = 0 
\ee
with the boundary condition
\be
{\cal G}_\Delta (x,y;0) = \delta^d(x-y) \; \; . 
\ee
The Green function of the operator $\Delta$ is then given by
\bea
G_\Delta(x,y) &=& \int_0^\infty {\cal G}_\Delta (x,y;t) \, dt 
\nonumber \\ 
\Delta (x) G_\Delta (x,y) &=& \delta^{d}(x-y) \; \; .
\eea
The singularities responsible for the divergences of the generating
functional arise from the short-distance behaviour ($y \rightarrow x$) 
of the Green function $G_\Delta(x,y)$. To extract these singularities,
we can use the asymptotic expansion of the heat kernel
for $t \rightarrow 0_+$:
\be
{\cal G}_\Delta (x,y;t) = \frac{1}{(4\pi t)^{d/2}} e^{-{|x-y|^2 \over 4t}}
  \sum_{n=0}^{\infty} a_n^\Delta (x,y) t^n \; \; .
\ee
The first Seeley-DeWitt coefficients are (in the 
coincidence limit $y \rightarrow x$) 
\bea
a_0^\Delta(x,x) &=& 1  \nonumber \\
a_1^\Delta(x,x) &=& -\sigma(x)  \nonumber \\
a_2^\Delta(x,x) &=& {1 \over 12} \gamma_{\mu \nu}(x) 
\gamma_{\mu \nu}(x) + {1 \over 2} \sigma^2(x)-{1 \over 6} d^2 \sigma(x) 
\eea
where $\gamma_{\mu \nu}$ is the usual field strength associated with 
$\gamma_\mu$. 

Correspondingly, we can write the Green function $G_\Delta$ as
\bea
G_\Delta(x,y) &=& G_0(x-y) a_0^\Delta(x,y)+ R_1(x-y;c\mu) a_1^\Delta(x,y)
\nonumber \\
&+& R_2(x-y;c\mu) a_2^\Delta(x,y) + \overline{G}_\Delta(x,y;c\mu) \; \; .
\label{G}
\eea
The various functions in (\ref{G}) are defined by \cite{JO}
\bea
G_0(x)&=&\frac{\Gamma(d/2-1)}{4\pi^{d/2}}|x|^{2-d}  \nonumber \\
R_1(x;c\mu)&=&\frac{1}{16 \pi^2} \left[ {-2 (c\mu)^{d-4}\over d-4}  +
  \frac{\Gamma (d/2-2)}{\pi^{d/2-2}} |x|^{4-d} \right]  \nonumber \\
R_2(x;c\mu)&=&\frac{|x|^2}{32 \pi^2} \left[ {(c\mu)^{d-4} \over d-4} +
  \frac{\Gamma (d/2-3)}{2 \pi^{d/2-2}} |x|^{4-d} \right] \; \; .
\label{propfunc}
\eea
We have regularized the short-distance  singularities by dimensional 
regularization. The scale-dependent coefficient functions $R_1$ and
$R_2$ are defined such that they are regular for 
$d \rightarrow 4$. The positive constant $c$ parametrizes different
regularization conventions ($c=1$ for minimal subtraction). We shall
follow the usual convention of CHPT \cite{GL84,GL85} with
\be
\ln c=-\frac{1}{2}\left[\ln 4\pi +\Gamma'(1)+1\right]
\label{lnc}
\ee
and employ the notation
\be
\Lambda=\frac{1}{16 \pi^2 (d-4)} \; \; . 
\label{Lambda}
\ee
By construction, $\overline{G}_\Delta(x,y;c\mu)$ has no poles in $d-4$
and is regular for $x \rightarrow y$, $d \to 4$ even for two derivatives. 
It is scale and convention dependent compensating the corresponding
dependences of $R_1$, $R_2$.

With these definitions, one obtains \cite{JO} the propagator and its 
derivatives in the coincidence limit $x \rightarrow y$ by 
analytic continuation in $d$:
\bea
G_\Delta(x,x) &=& -2 (c\mu)^{d-4}\Lambda \, a_1^\Delta (x,x) + 
\overline{G}_\Delta (x,x;c\mu)  \nonumber \\
d_\mu^xG_\Delta(x,y)_{|_{y=x}} &=& -2 (c\mu)^{d-4}\Lambda \,
 d^x_\mu a_1^\Delta (x,y)_{|_{y=x}} + d^x_\mu
\overline{G}_\Delta (x,y;c\mu)_{|_{y=x}}  \nonumber \\ 
d_\mu^x d_\nu^y G_\Delta(x,y)_{|_{y=x}} &=& -2 (c\mu)^{d-4}\Lambda
\left( d^x_\mu d^y_\nu  a_1^\Delta (x,y)_{|_{y=x}}+ 
{\delta_{\mu \nu} \over 2} a_2^\Delta (x,x) \right) \nn 
&& + d^x_\mu d^y_\nu \overline{G}_\Delta (x,y;c\mu)_{|_{y=x}} \; 
  \;  
\label{Gdiv}
\eea
where
\begin{equation} 
d_\mu^y G_\Delta(x,y)=\partial_\mu^y G_\Delta(x,y)-
G_\Delta(x,y) \gamma_\mu(y)~.
\end{equation} 

The one-loop singularities can also be obtained from the
heat-kernel representation as
\be
\mbox{Tr} \ln \left(\frac{\Delta}{\Delta_0}\right) = - \int_0^\infty
\frac{dt}{t} \int d^d x \left[ {\cal G}_\Delta(x,x;t)-{\cal
    G}_{\Delta_0}(x,x;t) \right] \; \; ,
\ee
leading to the well-known result
\be
\mbox{Tr} \ln \left(\frac{\Delta}{\Delta_0}\right) 
= 2 (c\mu)^{d-4}\Lambda\int d^d x \, \mbox{tr} \left[a_2^\Delta(x,x)-
a_2^{\Delta_0}(x,x) \right] + \mbox{finite terms} \; \; .
\label{logdet}
\ee

At the two-loop level the singularities come from the
short-distance behaviour of products of two or three propagators and
derivatives thereof (cf. Eq.~(\ref{z2})). To extract these 
singularities, one uses the representation (\ref{G}) and analyses
products of the three functions $G_0$, $R_1$, $R_2$ and their
derivatives at short distances. As an example,
\be
G_0(x)^2 \sim - \frac{1}{8 \pi^2(d-4)} \delta^d(x) \; \; .
\ee
All products relevant for our calculation were calculated first in
Ref.~\cite{JO}. We have checked their results and for convenience we
reproduce them in App.~\ref{app:JO}.  Each of the terms in Eq.~(\ref{z2})
has both local and nonlocal singularities. To be
able to renormalize the theory, the latter have to cancel in the sum. In
the following three subsections we will prove that this is indeed the
case. We first
consider the 1PR diagrams which have nonlocal singularities of the form
$A(x) G(x,y) B(y)$, and then the nonlocal subdivergences in the 1PI
diagrams. In the last subsection we will single out the divergences of the
form $1/(d-4) \times \ln m^2 $ which appear in all 1PI diagrams.
The requirement that they cancel leads to a nontrivial relation
between the coefficients of the double poles of the 1PI two-loop diagrams
(a and b in Fig.~\ref{fig:p6diag}) and the double poles of the 1PI one-loop
diagram (d in Fig.~\ref{fig:p6diag}).

\subsection{One-particle-reducible diagrams}

The three 1PR diagrams (c,e,f in 
Fig.~\ref{fig:p6diag}) are individually divergent and these
divergences are in general nonlocal. In a consistent quantum field
theory, these nonlocal divergences have to cancel. Only local
divergences (polynomials in masses and derivatives) are allowed
because it would otherwise be impossible to render the theory finite
by adding the local action $S_6$. 

The proof for this cancellation of nonlocal divergences runs as
follows. We can write the sum of the three 1PR diagrams as
\be
Z_2^{\mbox{\tiny 1PR}} = - {1 \over 2}
\left[S^{(1)}_{4\;i}+ {1 \over 2} S^{(3)}_{2\; ijk} G_{jk} \right]
G_{il}
\left[S^{(1)}_{4\;l}+ {1 \over 2} S^{(3)}_{2\; lrs} G_{rs} \right] \; \; .
\label{Z_1PR}
\ee
Recall that all quantities are to be taken at the classical field
$\phicl [{\bf j}]$. The terms inside the square brackets are nothing 
but the derivative of the $O(\hbar)$ term of the generating
functional, $Z_1$, which is finite by construction:
\be
S^{(1)}_{4\;i}+ {1 \over 2} S^{(3)}_{2\; ijk} G_{jk} = {
  \delta Z_1 \over \delta \phi_i} \; \; .
\ee
Since $Z_1$ is finite this would seem to prove that also 
$Z_2^{\mbox{\tiny 1PR}}$ in (\ref{Z_1PR}) is finite. The problem is
that the procedures of functional derivation and setting 
$\phi = \phicl [{\bf j}]$ do not commute in general because
$S_4$ may contain (divergent) terms that vanish at the classical field.
Denoting such terms as $S_4^{\mbox{\tiny EOM}}$, we may therefore have
\be
S_4^{\mbox{\tiny EOM}}[\phicl]=0, \; \; \; \; \; \;
\left( { \delta S_4^{\mbox{\tiny EOM}}[\phi] \over \delta
    \phi_i}\right)_{|_{\phi=\phicl}} \neq 0 \; \; ,
\ee
making $Z_2^{\mbox{\tiny 1PR}}$ divergent in general.
Such terms must be of the form
\be
S_4^{\mbox{\tiny EOM}}= \, \Xi \, X_i[\phi] { \delta S_2 \over \delta
  \phi_i}
\label{S4EOM} 
\ee
where $\, \Xi \,$ is a coupling constant that may be divergent for
$d \to 4$. 
Taking the functional derivative and setting $\phi=\phicl$, we obtain 
\be
\left({ \delta S_4^{\mbox{\tiny EOM}}[\phi] \over \delta
    \phi_j} \right)_{|_{\phi=\phicl}} = \, \Xi \, X_i[\phicl] 
S^{(2)}_{2 \; ij} =
\, \Xi \, X_i[\phicl] \Delta_{ij} \; \; .
\ee
Therefore, the contribution of such terms to the 1PR diagrams is
\be
-{1 \over 2} \,\Xi^2 \, X_i \Delta_{ij} X_j -  \, \Xi \, X_i { \delta Z^0_1
  \over \delta \phi_i} 
\; \; .
\label{Xi_1PR}
\ee 
Here, $Z^0_1$ is by definition free from terms that vanish at the
classical field and its derivative is therefore finite. The first term in
Eq.~(\ref{Xi_1PR}) is divergent (when $\Xi$ is divergent) but has the
form of a local action, which is perfectly allowed. On the other hand,
the second term is also divergent but contains a nonlocal piece
\be 
-{1 \over 2} \, \Xi \, X_i
S_{2\; ijk}^{(3)} \overline{G}_{jk} \; \; . 
\ee 
Such a nonlocal divergence must not be present in the final result.
In fact, it is rather easy to see that it is cancelled by the 1PI 
contribution of $S_4^{\mbox{\tiny EOM}}$ (diagram d in 
Fig.~\ref{fig:p6diag}): 
\bea
{1 \over 2} \left({ \delta^2 S_4^{\mbox{\tiny EOM}} \over \delta
\phi_j \delta
\phi_k}\right)_{|_{\phi=\phicl}} G_{jk} &=& 
{1 \over 2} \, \Xi \, \left[ {\delta X_i \over \delta
\phi_k} \Delta_{ij} + X_i S^{(3)}_{2\; ijk} \right] G_{jk} \nonumber \\ &=&
{1 \over 2} \, \Xi \, {\delta X_i \over \delta \phi_i} + {1 \over 2} \,
\Xi \, X_i S^{(3)}_{2\; ijk} G_{jk} \; \; .
\label{Xi_1PI}
\eea
The sum of (\ref{Xi_1PR}) and (\ref{Xi_1PI}) is in general
divergent but it has the form of a local action that can 
be renormalized by $S_6$.

\subsection{Absence of nonlocal subdivergences}
\label{sec:subdivs}
The three 1PI graphs (diagrams a,b,d in 
Fig.~\ref{fig:p6diag}) are the source of the
``true'' divergences of $O(\hbar^2)$. However, part of the
divergences occurring in the individual graphs have to cancel in the
sum. These are the so-called ``nonlocal subdivergences'' due to one-loop 
subgraphs. As is well known, these nonlocal subdivergences cancel if the
renormalization was properly done at $O(\hbar)$. Of course, we will 
verify this explicitly in our calculation (see also Ref.~\cite{BCE1}). 
Here we present a general proof within our formalism.

Let us start from the definition of the 1PI part of the 
generating functional of $O(\hbar^2)$: 
\be 
Z_2^{\mbox{\tiny 1PI}} = {1 \over 8}
S_{2\; ijkl}^{(4)} G_{ij} G_{kl} - {1 \over 12} S_{2\; ijk}^{(3)} G_{ir}
G_{js} G_{kt} S_{2\; rst}^{(3)} + {1 \over 2} S_{4\; ij}^{(2)} G_{ij} 
\; \; . \label{Z21PI}  
\ee 
The problematic divergences arise when one of the propagators in each
term in (\ref{Z21PI}) is replaced by its finite part $\overline{G}$:
\be 
Z_2^{\mbox{\tiny subdiv}} =
\left[ {1 \over 4} S_{2\; ijkl}^{(4)} G_{ij} - {1 \over 4} S_{2\;
ijk}^{(3)} G_{ir} G_{js} S_{2\; rsl}^{(3)} + {1 \over 2} S_{4\; kl}^{(2)}
\right] \overline{G}_{kl} \; \; .  
\ee 
As in the previous subsection, the key observation is that the term
between square brackets is proportional to the double functional derivative
of $Z_1$, the $O(\hbar)$ term in the generating functional, and
is therefore finite. Again this argument has a similar loophole as
before connected with the
possible presence of divergent terms that vanish at the classical
field. Because of such terms the finiteness of $Z_1$ 
(evaluated at the classical field) does not imply that its double functional
derivative is also finite. However, we have already seen that the nonlocal
subdivergences generated by such terms in the 1PI graphs are
compensated exactly by similar terms in the 1PR part.
This concludes the proof that nonlocal subdivergences cancel at
$O(\hbar^2)$.

\subsection{Weinberg consistency conditions}
\label{sec:Wein}

The fact that the residues of the poles in $d-4$ have to be 
polynomials in external momenta and masses \cite{collins} implies 
nontrivial relations between the coefficients of the
divergences of the three 1PI diagrams. To derive this consequence, we have
to look at the terms depending logarithmically on the masses that
appear in the residues of the single poles.
Diagrams a and b have the following structure:
\be
\mbox{`a+b'}= \int \!\! d^dx \mu^{2(d-4)} \left[ \Lambda^2
  \sum_a \alpha_a Y_a(x) + {\Lambda \over 16 \pi^2} \left(\sum_a
\beta_a Y_a(x) +M(x)
  \right) + {N(x) \over (16 \pi^2)^2} \right] \; \; ,
\label{eq:a+b}
\ee
where the $Y_a(x)$ are a complete set of local operators appearing in the 
action $S_6$, and $M(x)$, $N(x)$ are two generic functions which
contain the nonlocal terms generated by the loop integrals. By
definition, $M(x)$ does not contain any local terms that can be
expressed as linear combinations of the $Y_a(x)$.
The overall coefficient $\mu^{2(d-4)}$ ensures the correct
dimensions but is completely arbitrary, as in both diagrams the scale
$\mu$ does not appear at all. In fact, Eq.~(\ref{eq:a+b}) is
$\mu$-independent by definition, which implies the following equations:
\bea
\mu{\partial \alpha_a\over \partial \mu}&=&0  \nn
\label{eq:abrge}
\mu{\partial \beta_a\over \partial \mu} &=&
-2  \alpha_a \\
\mu{\partial M(x)\over \partial \mu} &=& 0 \nn
\mu{\partial N(x) \over \partial \mu}&=&-2\left(\sum_a \beta_a 
Y_a(x) +M(x) \right) \; \; . \nonumber
\eea
From Eq.~(\ref{eq:abrge}) it is clear that the residue of the single pole
in $d-4$ contains terms of the form $\ln m^2/\mu^2$, where $m$ is one of
the masses of the particles in the theory. Such terms can only be 
cancelled by analogous ones in the one-loop diagram d. The latter is of 
the form
\be
\mbox{`d'}= \int \!\! d^dx \sum_i \hat L_i \mu^{d-4} \left(
  \Lambda \sum_a \eta_a^i Y_a(x) + {1 \over 16 \pi^2} H_i(x) \right) \; \; ,
\ee
where the $\hat L_i$ are the coefficients\footnote{We anticipate here 
the notation that we
  will use in the CHPT case. The argument, however, remains completely
  general.}  in front of the operators $X_i$ appearing 
in the action $S_4$: $S_4=\int d^dx \sum_i \hat L_i X_i$. $H_i(x)$ are
generic functions containing the nonlocal part of the loop integrals.
For each $i$, the one-loop diagram is again $\mu$-independent. Since
all the $\hat L_i$ are themselves $\mu$-independent by definition we obtain
\bea
\mu{\partial \eta_a^i\over \partial \mu}&=&0  \nn
\mu{\partial H_i(x) \over \partial \mu}&=&-\sum_a \eta_a^i Y_a(x) \; \; .
\eea
The $\hat L_i$ are in general divergent,
\be
\hat L_i = \mu^{d-4} \left({\hat \Gamma}_i \Lambda +\hat L_i^r(\mu,d) \right) \;
\; ,
\ee
and therefore also the one-loop diagram has logarithms of the masses in the 
residue of the single pole. The condition that these logarithms cancel in
the sum implies the following relations \cite{Wein79}:
\be 
\alpha_a = - {1 \over 2} \sum_i {\hat \Gamma}_i \eta_a^i \label{eq:WR} ~.
\ee 
The cancellation of the other nonlocal terms requires
\be
M(x)=-\sum_i {\hat \Gamma}_i \bar H_i(x) \; \; ,
\label{nonlocaldivs}
\ee
where
\be
\bar H_i(x)= H_i(x) - {1\over 2} \ln {m^2 \over \mu^2} \sum_a \eta^i_a
Y_a(x) \; \; .
\ee

\setcounter{equation}{0}
\section{Chiral perturbation theory to two loops}
\label{sec:CHPT}

Having set up the general framework, we now specialize to CHPT. As we will
see, the divergence calculation is now only a matter of (long and tedious)
algebraic manipulations. The procedure, on the other hand, is well defined
and completely straightforward. It amounts to the following steps:
\begin{enumerate}
\item 
Define the first three terms in the action: $S_2$, $S_4$ and $S_6$;
\item
Define the differential operator $\Delta$ for CHPT and calculate the
corresponding Seeley-DeWitt coefficients;
\item
Expand $S_2$ ($S_4$) up to fourth (second) order in $\xi$;
\item
Contract indices between the vertices $S_{2n\, i_1,\ldots
  i_m}^{(m)}$ and the Seeley-DeWitt coefficients appropriately;
\item 
Finally, reduce everything to the standard basis defined by $S_6$.
\end{enumerate}

In the following we will give all the necessary ingredients for this
lengthy calculation. In the next section, we present our final results 
for chiral $SU(n)$ and for the physically relevant cases $n=2,3$. 

\subsection{Preliminaries}

In Euclidean space, the $O(p^2)$ Lagrangian is given by \cite{GL84,GL85}
\be
{\cal L}_2 = {F^2 \over 4} \lgl u_\mu u_\mu - \chi_+ \rgl 
\ee
with
\begin{eqnarray} 
u_\mu &=& i \{ u^\dg(\partial_\mu - i r_\mu)u - 
u(\partial_\mu - i \ell_\mu) u^\dg\} \nn
\chi_\pm &=&  u^\dg \chi u^\dg \pm u \chi^\dg u\,.  
\end{eqnarray}
Here, $u(\phi)$ is the usual matrix field of Goldstone bosons and the
external fields $v_\mu,a_\mu,s,p$ are contained in  $r_\mu = v_\mu + a_\mu$, 
$\ell_\mu = v_\mu - a_\mu$, $\chi= 2 B (s+ip)$. The symbol
$\lgl \dots \rgl$ denotes the $n$-dimensional flavour trace for chiral
$SU(n)$ and $F,B$ are the two low-energy constants of $O(p^2)$. The
EOM (\ref{EOM}) take the form
\be
\nabla_\mu u_\mu + {i \over 2} \hat\chi_- = 0 
\label{EOMCHPT}
\ee
with
\begin{eqnarray} 
\nabla_\mu X &=& \partial_\mu X + [\Gamma_\mu,X]\nn
\Gamma_\mu &=& {1 \over 2} \left[ u^\dagger ( \partial_\mu - ir_\mu) u +
  u( \partial_\mu -il_\mu) u^\dagger \right]\nn
\hat\chi_- &=& \chi_--1/n\lgl \chi_- \rgl~.
\end{eqnarray} 

Expanding ${\cal L}_2$ around the classical field, the solution 
of the EOM (\ref{EOMCHPT}), we obtain 
the $(n^2-1)$-dimensional matrix-differential operator $\Delta$:
\be
\Delta = -d_\mu d_\mu + \sigma \; \; ,
\ee
where
\bea
d_\mu &=& \partial_\mu + \gamma_\mu  \nn
\gamma_{\mu \, a b} &=& -{1 \over 2} \lgl \Gamma_\mu [ \lambda_a, \lambda_b ]
\rgl  \nn
\sigma_{ab} &=& {1 \over 8} \lgl [u_\mu, \lambda_a][u_\mu, \lambda_b] +
\left\{ \lambda_a , \lambda_b \right\} \chi_+ \rgl \; \; .
\eea
We use the normalization $\lgl \lambda_a \lambda_b \rgl = 2
\delta_{ab}$ and we define the fluctuation fields $\xi$ via
\begin{equation} 
u^2(\phi) = u(\phicl) e^{\frac{i\sqrt{2}}{F}\xi} u(\phicl)~.
\end{equation} 

In the notation of the previous section, the third- and fourth-order
terms in the expansion around the classical field are given by:
\bea
S_2^{(3)} \xi^3 &=& -{\sqrt{2} \over F} \int d^dx \left[ 3 \lgl \xi \nabla_\mu \xi 
  \xi u_\mu \rgl +{i \over 2n} \lgl \chi_- \rgl \lgl \xi^3 \rgl \right] 
\label{S2_3}\\ 
S_2^{(4)} \xi^4 &=& {2 \over F^2} \int d^dx \lgl [ \xi, \nabla_\mu \xi ] [ \xi,
\nabla_\mu \xi ] +{1 \over 4} \left[ \xi , [ u_\mu, \xi] \right]^2 - {1
  \over 2} \chi_+ \xi^4 \rgl \; \; .
\label{S2_4}
\eea

The most general chiral $SU(n)$  Lagrangian of $O(p^4)$ is given by
\bea
{\cal L}_4 &=& \sum_{i=0}^{12} {\hat L}_i X_i + \mbox{contact terms}\nn  
&=& -{\hat L}_0 \lgl u_\mu u_\nu u_\mu u_\nu \rgl -{\hat L}_1 \lgl  u\ccdot u \rgl^2 
- {\hat L}_2 \lgl u_\mu u_\nu \rgl \lgl u_\mu u_\nu \rgl
- {\hat L}_3 \lgl ( u\ccdot u )^2 \rgl \nn
&+& {\hat L}_4 \lgl u\ccdot u \rgl \lgl \chi_+
\rgl + {\hat L}_5 \lgl u\ccdot u \chi_+ \rgl 
- {\hat L}_6 \lgl \chi_+ \rgl^2 - {\hat L}_7 \lgl \chi_- \rgl^2  \nonumber \\
&-& {{\hat L}_8 \over 2} \lgl \chi_+^2 + \chi_-^2 \rgl + i{\hat L}_9 \lgl
f_{+ \,\mu \nu} u_\mu u_\nu \rgl - {{\hat L}_{10} \over 4} 
\lgl f_+^2 - f_-^2 \rgl \nn
&+& i{\hat L}_{11} \lgl \hat\chi_-( \nabla_\mu u_\mu + {i \over 2}
\hat\chi_- ) \rgl - {\hat L}_{12} \lgl ( \nabla_\mu u_\mu + {i \over 2}
\hat\chi_- )^2 \rgl \nn 
&+& \mbox{contact terms} \; \; , 
\label{L4N}
\eea
where $u\ccdot u = u_\mu u_\mu$, $ f_\pm^2=f_{\pm \, \mu \nu} f_{\pm \; \mu
  \nu}$ and 
\begin{eqnarray} 
f_{\pm \; \mu\nu} &=& u F_{L \; \mu\nu} u^\dg \pm u^\dg F_{R \;
\mu\nu} u  \nn
F_{R \; \mu\nu} &=& \partial_\mu r_\nu - \partial_\nu r_\mu -
i[r_\mu,r_\nu] \nn
F_{L \; \mu\nu} &=& \partial_\mu \ell_\nu - \partial_\nu \ell_\mu -
i [\ell_\mu,\ell_\nu]~.
\end{eqnarray}

The renormalization at the one-loop level is performed by splitting the
constants ${\hat L}_i$ into a divergent and a finite part:
\be
{\hat L}_i=(c\mu)^{d-4}\left[{\hat \Gamma}_i \Lambda + {\hat L}_i^r(\mu,d) 
\right]\; \; ,
\label{Lhatren}
\ee
where $c$ and $\Lambda$ are defined in Eqs.~(\ref{lnc}), 
(\ref{Lambda}).
The measurable low-energy constants ${\hat L}_i^r(\mu)$ (or rather
the corresponding constants for $n=2$ or $3$) are given by
\begin{equation}  
{\hat L}_i^r(\mu) = {\hat L}_i^r(\mu,4)~.
\label{Lhatphys}
\end{equation}   
The expressions for the coefficients ${\hat \Gamma}_i$ can be obtained from
Eq.~(\ref{logdet}) by taking the trace of the Seeley-DeWitt coefficient
$a_2(x,x)$ in CHPT \cite{GL85}:
\bea
{\hat \Gamma}_0={n \over 48}\; , & &{\hat \Gamma}_1={1 \over 16}\; ,\qquad 
{\hat \Gamma}_2 ={1 \over 8}\; , \nn
{\hat \Gamma}_3={n \over 24}\; , & &{\hat \Gamma}_4={1 \over 8}\; , \qquad 
{\hat \Gamma}_5={n \over 8}\; , \nn
{\hat \Gamma}_6={n^2+2 \over 16 n^2}\; , & &{\hat \Gamma}_7=0\; , \qquad
{\hat \Gamma}_8={n^2-4 \over 16 n}\; , \nn
{\hat \Gamma}_9={n \over 12}\; , & &{\hat \Gamma}_{10}=-{n \over 12}\; \; .
\eea

For $n=2,3$, not all the monomials in the Lagrangian (\ref{L4N}) are
linearly independent. The $SU(3)$ Lagrangian of Ref.~\cite{GL85} is
recovered with the help of the relations
\begin{eqnarray} 
L_1={\hat L}_0/2+ {\hat L}_1 \; , \qquad L_2={\hat L}_0+ {\hat L}_2 
\; , \qquad L_3=-2 {\hat L}_0+ {\hat L}_3 \nn
L_i={\hat L}_i \; (i=4, \ldots, 10) \; , \qquad
{\hat L}_{11}={\hat L}_{12}=0 \; .
\label{LiSU3}
\end{eqnarray}
The corresponding relations for the $SU(2)$ Lagrangian of
Ref.~\cite{GL84} are
\begin{eqnarray} 
l_1=-2{\hat L}_0 + 4 {\hat L}_1 + 2 {\hat L}_3 \; , \qquad
l_2=4({\hat L}_0 + {\hat L}_2) \nn
l_3=4(-2{\hat L}_4 - {\hat L}_5 + 4 {\hat L}_6 + 2 {\hat L}_8) \; ,
\qquad l_4=4(2{\hat L}_4 + {\hat L}_5)\nn
l_5={\hat L}_{10} \; , \quad l_6= -2 {\hat L}_9 \; , \quad
l_7=-8(2 {\hat L}_7 + {\hat L}_8)\nn
{\hat L}_{11}= - l_4/4 \; , \qquad {\hat L}_{12}= 0 ~.
\label{liSU2}
\end{eqnarray} 
Note that there is an EOM term of the type 
(\ref{S4EOM}) proportional to $l_4$ in the two-flavour case. 

\subsection{One-loop diagram}

To calculate the divergences arising from the one-loop diagram with 
${\cal L}_4$ (diagram d in Fig.~\ref{fig:p6diag}),
we first need to expand the thirteen terms in the 
Lagrangian (\ref{L4N}) in the fluctuation fields $\xi$ according to
\be
\label{eq:xi_exp}
X_i = X^{(0)}_i+X_i^{(1)}\xi+{1 \over 2}X_i^{(2)}\xi^2 + O(\xi^3)\,,
\ee 
where the coefficients $X^{(0)}_i$, $X_i^{(1)}$ and $X_i^{(2)}$ are to
be taken at the classical field. Their explicit expressions are as 
follows:
\bea
X_0 &=& X^{(0)}_0 +\frac{4\sqrt{2}}{F} \lgl \nabla_\mu \xi u_\nu u_\mu 
u_\nu \rgl - \frac{1}{F^2}\left(8 \lgl \nabla_\mu \xi \nabla_\nu \xi 
u_\mu u_\nu \rgl + 4 \lgl \nabla_\mu \xi u_\nu \nabla_\mu \xi u_\nu 
\rgl \right. \nn
&& \left. +  \lgl u_\mu u_\nu u_\mu \left[ \xi, \left[ 
u_\nu, \xi \right] \right] \rgl \right) + O(\xi^3)   \nn
X_1 &=& X^{(0)}_1 + \frac{4\sqrt{2}}{F} \lgl u\ccdot u \rgl \lgl 
\nabla_\nu \xi u_\nu \rgl - \frac{8}{F^2} 
\lgl \nabla_\mu \xi u_\mu \rgl \lgl \nabla_\nu \xi u_\nu \rgl \nn
&& - \frac{4}{F^2} \lgl u\ccdot u \rgl \left( \lgl \nabla_\nu \xi 
\nabla_\nu \xi \rgl
  + {1 \over 4} \lgl \left[ u_\nu, \xi \right] \left[ u_\nu , \xi \right]
  \rgl \right) + O(\xi^3)  \nn
X_2 &=& X^{(0)}_2 + \frac{4\sqrt{2}}{F} \lgl u_\mu u_\nu \rgl \lgl
u_\mu \nabla_\nu \xi \rgl - \frac{4}{F^2}
\lgl u_\mu \nabla_\nu \xi \rgl \lgl u_\mu \nabla_\nu \xi + u_\nu
\nabla_\mu \xi \rgl \nn 
&& - \frac{4}{F^2} \lgl u_\mu u_\nu \rgl \left( \lgl \nabla_\mu \xi 
\nabla_\nu \xi \rgl 
  + {1 \over 4} \lgl \left[u_\mu , \xi \right] \left[ u_\nu , \xi \right]
  \rgl \right) +O(\xi^3)  \nn
X_3 &=& X^{(0)}_3+\frac{2\sqrt{2}}{F} \lgl u\ccdot u \left\{ u_\nu , 
\nabla_\nu \xi \right\} \rgl - \frac{1}{F^2}\left(2\lgl \left\{ u_\mu , 
\nabla_\mu \xi \right\} \left\{ u_\nu
  , \nabla_\nu \xi \right\} \rgl\right. \nn 
&&\left.+4 \lgl u\ccdot u \nabla_\nu \xi \nabla_\nu \xi \rgl +
\lgl \left\{ u\ccdot u , u_\nu \right\} \xi \left[ u_\nu  
  , \xi \right] \rgl\right) +O(\xi^3)  \nn
X_4 &=& X^{(0)}_4-\frac{2\sqrt{2}}{F} \lgl u_\mu \nabla_\mu \xi \rgl 
\lgl \chi_+ \rgl - \frac{i\sqrt{2}}{F}
\lgl u\ccdot u \rgl \lgl \xi \chi_- \rgl \nn 
&& - \frac{1}{F^2}\left( \lgl u\ccdot u \rgl \lgl \xi^2 \chi_+ \rgl 
 -4 i \lgl u_\mu \nabla_\mu \xi \rgl \lgl \xi \chi_- \rgl\right. \nn 
&& \left.-2 \lgl \chi_+ \rgl \lgl \nabla_\mu \xi \nabla_\mu \xi + {1 \over 4}
\left[u_\mu, \xi \right] \left[ u_\mu, \xi \right] \rgl\right) +O(\xi^3) 
\nn 
X_5 &=& X^{(0)}_5-\frac{i\sqrt{2}}{2 F} \lgl u\ccdot u \left\{ \xi,
\chi_- \right\}
\rgl -\frac{\sqrt{2}}{F} \lgl \chi_+ \left\{ u_\mu, \nabla_\mu \xi 
\right\} \rgl \nn
&& + \frac{1}{F^2}\left(2\lgl \chi_+ \nabla_\mu \xi \nabla_\mu \xi 
\rgl + i \lgl
\left\{ \xi, \chi_- \right\} \left\{ u_\mu, \nabla_\mu \xi \right\} \rgl
\right.\nn
&& \left.+ {1 \over 2} \lgl \chi_+ \left[ u_\mu, \xi \right] \left[ u_\mu , \xi 
\right] \rgl -{1 \over 2} \lgl \xi^2 \left\{ \chi_+, u\ccdot u \right\}
\rgl\right) +O(\xi^3) \nn
X_6 &=& X^{(0)}_6 +\frac{2i\sqrt{2}}{F}\lgl \chi_+ \rgl \lgl \xi
\chi_- \rgl +\frac{2}{F^2}\left( 
\lgl \xi \chi_- \rgl^2 + \lgl \chi_+ \rgl \lgl \xi^2 \chi_+ 
\rgl  \right)+O(\xi^3) \nn
X_7 &=& X^{(0)}_7 +\frac{2i\sqrt{2}}{F} \lgl \chi_- \rgl \lgl \xi 
\chi_+ \rgl +\frac{2}{F^2}\left( 
\lgl \xi \chi_+ \rgl^2 + \lgl \chi_- \rgl \lgl \xi^2 \chi_- 
\rgl \right) +O(\xi^3) \nn
X_8 &=& X^{(0)}_8 + \frac{i\sqrt{2}}{F} \lgl \xi \left\{ \chi_+ , 
\chi_- \right\} \rgl + \frac{1}{2F^2}
 \lgl \left\{ \xi , \chi_+ \right\} \left\{ \xi, \chi_+ \right\} 
+ \left\{ \xi , \chi_- \right\}\left\{ \xi , \chi_- \right\} \rgl 
+O(\xi^3) \nn
X_9 &=&X^{(0)}_9-\frac{i\sqrt{2}}{F} \lgl f_{+ \, \mu \nu} 
\left[ u_\mu , \nabla_\nu
  \xi \right] \rgl -\frac{\sqrt{2}}{2F} \lgl \xi \left[ f_{- \, \mu
\nu} , u_\mu
  u_\nu \right] \rgl \nn
&& + {i \over 2 F^2} \lgl f_{+ \, \mu \nu} \left( \left[ u_\mu, \xi\right]
  \left[ u_\nu , \xi \right] + \left[ \xi, \left[ u_\mu u_\nu , \xi
    \right] \right] \right) \rgl \nn
&& + \frac{1}{F^2}\left(2i \lgl f_{+\, \mu \nu} \nabla_\mu \xi 
\nabla_\nu \xi \rgl - 
 \lgl \left[ f_{- \, \mu \nu} , \xi \right] \left[ \nabla_\mu \xi,
  u_\nu \right] \rgl\right) +O(\xi^3)  \nn  
X_{10} &=& X^{(0)}_{10} - \frac{i\sqrt{2}}{2F} \lgl \xi \left[ 
f_{- \, \mu \nu},
  f_{+ \, \mu \nu} \right] \rgl \nn
&& -{1 \over 2 F^2} \lgl \xi \left( f_{- \, \mu \nu} \left[ f_{- \, \mu \nu},
    \xi \right] - f_{+ \, \mu \nu} \left[ f_{+ \, \mu \nu},
    \xi \right] \right) \rgl +O(\xi^3)  \nn
X_{11} &=&  \frac{i\sqrt{2}}{F} \lgl \hat\chi_- \left( - \nabla^2 \xi +{1 \over 
    4} \left[ u_\mu, \left[\xi, u_\mu \right] \right] + {1 \over 4} \left\{ 
    \xi , \chi_+ \right\} \right) \rgl \nn
&& + \frac{1}{F^2}\left(i \lgl \left[ \hat\chi_- , \nabla_\mu \xi \right] \left[
  u_\mu, \xi \right] \rgl + {1 \over 2} \lgl \hat\chi_- \xi \chi_- \xi 
\rgl \right) +O(\xi^3 ) \nn
X_{12} &=& - \frac{2}{F^2}\Delta \xi \Delta \xi+O(\xi^3) \; \; .
\eea
We recall that $X^{(0)}_{11}=X^{(0)}_{12}=0$ due to the EOM
(\ref{EOMCHPT}). In fact, $X_{12}$ can be disposed of altogether
\cite{BCE1}. Because $X_{12}^{(1)}=0$, ${\hat L}_{12}$ cannot contribute to
the 1PR diagrams e,f in Fig.~\ref{fig:p6diag}. Moreover, the
one-loop diagram d
proportional to  ${\hat L}_{12}$ vanishes in dimensional regularization
because of $\delta^d(0)=0$. We can therefore set ${\hat L}_{12}=0$ without
loss of generality. As a consequence, $S_4^{\mbox{\tiny EOM}}$ in
(\ref{S4EOM}) depends on a single parameter ${\hat L}_{11}$ in CHPT.

The calculation of the divergences of the one-loop diagram d in 
Fig.~\ref{fig:p6diag} is then rather straightforward:
we just need to contract the two $\xi$'s in the terms of $O(\xi^2)$
into the propagator $G$ and use the relations (\ref{Gdiv}).
For illustration, let us consider one simple case, the operator $X_6$.
According to the definition (\ref{eq:xi_exp}) we have
\be
\label{eq:X6_2}
{1 \over 2} X_{6 \; ab}^{(2)} = {1 \over 2 F^2} \left[ \lgl \chi_+ \rgl \lgl
  \chi_+ \{ \lambda_a, \lambda_b \} \rgl + 2 \lgl \chi_- \lambda_a \rgl
  \lgl \chi_- \lambda_b \rgl \right] \; \; ,
\ee
where we have expanded $\xi$ in components as $\xi = 1/\sqrt{2} \lambda_a
\xi_a$. According to (\ref{Gdiv}), the divergent part generated by $X_6$ in 
the one-loop graph is proportional to $X_{6 \; ab}^{(2)} \sigma_{ab}$. To
contract the indices, we  use the completeness relations for $SU(n)$:
\bea
\sum_{a=1}^{n^2-1} \lgl \lambda_a A \lambda_a B \rgl = -{2 \over n} \lgl AB 
\rgl + 2 \lgl A \rgl \lgl B \rgl  \nonumber \\
\sum_{a=1}^{n^2-1} \lgl \lambda_a A \rgl \lgl \lambda_a B \rgl = 2 \lgl AB  
\rgl -{2 \over n} \lgl A \rgl \lgl B \rgl \; \; .
\label{eq:complr}
\eea
The final result reads
\bea
Z_2^{\mbox{\tiny{div}}}({\hat L}_6) &=& 
{\hat L}_6 {(c\mu)^{d-4} \over F^2} \Lambda \! \int \! \! d^dx \left\{-n \lgl 
  \chi_+ \rgl \lgl \chi_+ u\ccdot u \rgl + \left(n -{4\over n}\right)
  \lgl \chi_+^2 \rgl \lgl \chi_+ \rgl  \right. \nonumber \\
&-& \lgl \chi_+ \rgl^2 \lgl u\ccdot u 
  \rgl +\left(1 +{2 \over n^2 }\right) \lgl \chi_+ \rgl^3 -2 \lgl \chi_-^2
u\ccdot u \rgl + 2 \lgl \chi_- u_\mu \chi_- u_\mu \rgl \nonumber \\ 
&+& \left. 2 \lgl \chi_+ \chi_-^2 \rgl - {4 \over n} \lgl \chi_+ \chi_-
  \rgl \lgl \chi_- \rgl + {2 \over n^2} \lgl \chi_+ \rgl \lgl \chi_- \rgl^2 
  \right\} \; \; .
\eea

\subsection{Butterfly diagram}

To calculate the butterfly diagram a in Fig.~\ref{fig:p6diag}, the 
expansion of $S_2$ is needed 
to fourth order in $\xi$ as given in
Eq.~(\ref{S2_4}). The divergent part of this diagram can again be obtained
from the relations (\ref{Gdiv}), since the diagram is the product of two
one-loop graphs. To illustrate how the calculation is performed, we again
single out a simple piece in the four-field vertex $S^{(4)}_2$, namely
\be
B^{(4)}_{abcd} = -{ 1 \over 96 F^2} \sum_{\mbox{\tiny{perm.}}} \lgl \chi_+
\lambda_a  \lambda_b \lambda_c \lambda_d \rgl \; \; , 
\ee
where the sum goes over all possible permutations of the four 
indices $abcd$. The two-loop divergence is proportional to $B^{(4)}_{abcd}
\sigma_{ab} \sigma_{cd}$. The calculation is again rather simple,
requiring only an iterated use of the completeness relations
(\ref{eq:complr}). The result reads
\bea
Z_2^{\mbox{\tiny{div}}}(B^{(4)}) \!&=&\! \!
-\frac{(c\mu)^{2(d-4)}\Lambda^2}{48 F^2}\! \int \! \! d^dx \left\{\!
{ n^2 \over 2} \lgl \chi_+ (u\ccdot u)^2 \rgl-2 \lgl \chi_+ u_\mu u\ccdot u
u_\mu \rgl  \right. \nonumber \\
&+& 2 \lgl \chi_+ u_\mu u_\nu u_\mu u_\nu \rgl+{n \over 2} \lgl \chi_+ \rgl
\lgl (u\ccdot u)^2 \rgl+2n \lgl \chi_+ u\ccdot u \rgl \lgl u\ccdot u \rgl
 \nonumber \\
&-& n \lgl \chi_+ u_\mu \rgl \lgl u_\mu u\ccdot u\rgl +\lgl \chi_+ \rgl
\lgl u\ccdot u \rgl^2 + (6-n^2) \lgl \chi_+^2 u\ccdot u 
\rgl+2 \lgl (\chi_+ u_\mu)^2 \rgl  \nonumber \\
&-& 2\left(n-\frac{4}{n} \right) \lgl
\chi_+^2 \rgl \lgl u\ccdot u \rgl-\left( 3n -\frac{2}{n} \right) \lgl \chi_+
\rgl \lgl \chi_+ u\ccdot u \rgl 
 \nonumber \\
&+& \left(n-\frac{8}{n}\right) \lgl \chi_+ u_\mu \rgl^2- 2 \left(1+{3 \over
    n^2} \right) \lgl \chi_+ \rgl^2 \lgl u\ccdot u  
  \rgl   \nonumber \\
&+& \left(\frac{n^2}{2}-6+\frac{24}{n^2} \right) \lgl \chi_+ ^3 \rgl+
 n\left( {5 \over 2}-{10 \over n^2} -{24 \over n^4} \right) \lgl 
  \chi_+^2 \rgl \lgl 
\chi_+ \rgl \nonumber \\
&+& \left. \left( 1 +{6 \over n^2}+{6 \over n^4} \right) \lgl \chi_+\rgl^3
\right\} \; \; . 
\eea
The term written down here
is only the one containing a double pole in $d-4$. The same diagram with the 
$B^{(4)}$ vertex does also contain a single pole which is however
nonlocal, being proportional to $\Gb$. As discussed in the previous
section, this term drops out in the final
result. The calculation of the rest of the diagram is algebraically more
involved but conceptually analogous to what we have shown here.

\subsection{Sunset diagram}

For the sunset diagram b in Fig.~\ref{fig:p6diag} we need the term of 
order $\xi^3$ in the expansion
of $S_2$ as given in Eq.~(\ref{S2_3}).
To calculate the divergent part of this diagram is the most difficult 
part of the whole calculation. In fact, it is the only genuine two-loop
diagram occurring in this calculation.
Its treatment requires the use of all the relations
given in App.~\ref{app:JO} for the products of two or three singular functions
$G_0$, $R_1$ and $R_2$ and derivatives thereof \cite{JO}.
To illustrate how the calculation proceeds, we again isolate a simple part 
of the whole diagram, i.e., we take for the three-field vertex the part
\be
B^{(3)}_{abc} = -\frac{i}{8 n F} \lgl \chi_- \rgl \lgl \lambda_a
\{\lambda_b, \lambda_c\} \rgl \; \; .
\ee
The divergent part generated by this vertex entering the sunset diagram is
proportional to
\be
 \int d^dx d^dy \lgl \chi_-(x) \rgl 
\lgl \lambda_a \lambda_b \lambda_c \rgl
G_{ad}(x,y) G_{be}(x,y) G_{cf}(x,y) 
\lgl \lambda_d \lambda_e \lambda_f \rgl
\lgl \chi_-(y) \rgl \; \; .
\ee
The product of three propagators without derivatives diverges at short
distances. By using the list of equations in App.~\ref{app:JO} one gets
\bea
&&G_{ad}(x,y) G_{be}(x,y) G_{cf}(x,y) =(c\mu)^{2(d-4)}\Lambda \times \nn
&& \left[-{1 \over 32
\pi^2} \partial_x^2 \delta^d(z) a_{0\; ad}(x,y) a_{0\; be}(x,y)
  a_{0\; cf}(x,y) \right. + \nn
&&\left. \left( 2 \Lambda - \frac{1}{16 \pi^2} \right) 
    \delta^d(z) \left( \delta_{ad} \delta_{be} a_{1\; cf}(x,x)+\delta_{ad}
      a_{1\; be}(x,x) \delta_{cf} 
 +  a_{1\; ad}(x,x)  \delta_{be} \delta_{cf} \right) \right] \nn
&&+ \ldots \; \;  ,
\label{eq:G^3}
\eea
where $z=x-y$ and we have used $a_{0\; ab}(x,x)=\delta_{ab}$.
The ellipsis stands for terms that do not contribute
local divergences to the generating functional. In fact, the terms not
written contain a divergent nonlocal piece proportional to $G_0^2$ which
drops out in the final result. After removing the derivative from
the $\delta$-function by partial integration, one uses again the 
completeness relations (\ref{eq:complr}) to obtain the final result
\bea
Z_2^{\mbox{\tiny{div}}}(B^{(3)})&=& -{(c\mu)^{2(d-4)} \over 48 F^2} \Lambda
 \int d^dx \left\{ 3 \left(\Lambda - {1 \over 32 \pi^2} \right)
 \right. \times \nn   
& & \left[ \left(1-{5 \over n^2} +{4 \over n^4} \right) \lgl \chi_+ \rgl \lgl
  \chi_- \rgl^2
- \left(1-\frac{4}{n^2} \right) \lgl \chi_- \rgl^2 \lgl u\ccdot u \rgl 
 \right] \nn
&&+ \left.{n \over 32 \pi^2} \left(1-{5 \over n^2} +{4 \over n^4} \right) 
   \lgl \chi_- \rgl \lgl \nabla^2 \chi_- \rgl \right\}
\; \; .
\eea
The rest of the sunset diagram is much more involved and produces many
more terms but is conceptually completely analogous to the piece we have
treated here.

\subsection{Checks}

In such a long calculation it is mandatory to make all kinds of
possible checks during the calculation and on the final result. 
Our calculation has successfully passed the following tests:
\begin{enumerate}
\item
We have explicitly verified the 115 Weinberg relations (\ref{eq:WR})
for $SU(n)$.
\item
We have explicitly verified that all the nonlocal divergences appearing
at intermediate stages of the calculation cancel in the final
result. This cancellation corresponds to Eq.~(\ref{nonlocaldivs}) and is
described in App.~\ref{app:subdivs}.
\item
From our final results for $n=2$ and $3$ we have extracted the divergences
appearing in most of the processes that have already been calculated
to two-loop accuracy. The comparison with results available in the
literature is performed in Sec.~\ref{sec:app}.
\end{enumerate}
Moreover, all parts of the calculation were done independently by at
least two of us. 

\setcounter{equation}{0}
\section{Results}
\label{sec:results}

With the formulas of the previous sections we can extract all
divergent parts of the generating functional. We then use the methods
of Ref.~\cite{BCE2} to convert them into the minimal number of terms.

The infinities of the effective action are then given by the generating
functional
\begin{equation}
z^{(n){\rm inf}}_2 = \frac{(c\mu)^{2(d-4)}}{F^2}\sum_{i=1}^{115} 
\left(\hat{\Gamma}^{(2)}_i\Lambda^2 + 
(\hat{\Gamma}^{(1)}_i+\hat{\Gamma}^{(L)}_i(\mu,d))\Lambda\right)\int d^d x
\,Y_i 
\end{equation}
for the case of $n$ flavours. The structures $Y_i$ were derived in
Ref.~\cite{BCE2} and are given
in Table \ref{tab:nf1}. The coefficients $\hat{\Gamma}^{(2)}_i$
and $\hat{\Gamma}^{(1)}_i$
can be found in Table \ref{tab:nf1} and $\hat{\Gamma}^{(L)}_i$ in Table
\ref{tab:nf2}. The $\hat{\Gamma}^{(L)}_i$ are linear combinations of the 
renormalized coupling constants ${\hat L}_i^r(\mu,d)$ of $O(p^4)$
defined in Eq.~(\ref{Lhatren}). We have set ${\hat L}_{11}=0$ for
simplicity. Moreover, we have converted the infinities to Minkowski
conventions since most diagrammatic
calculations are performed in Minkowski space.
In practice, this means that the coefficients of the terms with 0 or 4
derivatives or vector, axial-vector external fields change
sign while those with 2 or 6 remain the same.

In order to subtract the infinities and obtain a finite generating
functional, it then suffices to replace in the chiral Lagrangian of
$O(p^6)$,
\begin{equation}
{\cal L}_6 = \sum_{i=1}^{115} K_i Y_i ~,
\end{equation}
the constants $K_i$ by
\begin{equation}
K_i = {(c\mu)^{2(d-4)} \over F^2} \left(
  K_i^r(\mu,d)-\hat{\Gamma}^{(2)}_i\Lambda^2 
  -(\hat{\Gamma}^{(1)}_i+\hat{\Gamma}^{(L)}_i(\mu,d))\Lambda\right)\,.
\end{equation}
Of course, all physical quantities are
defined for $d=4$, e.g., $K_i^r(\mu):= K_i^r(\mu,4)$ as in 
Eq.~(\ref{Lhatphys}). 
We recall that $c = 1$ corresponds to minimal subtraction
while the conventional version of modified minimal subtraction used 
in CHPT corresponds to
\begin{equation}
\ln c = -\frac{1}{2}\left(\ln 4\pi +\Gamma^\prime(1)+1\right)\,.
\end{equation}

The scale independence of the constants 
${\hat L}_i$ and $K_i$ implies the following renormalization group 
equations for the renormalized low-energy constants (see also 
Ref.~\cite{BCEGS2}):
\begin{equation} 
\mu\frac{d K_i^r(\mu)}{d\mu}=
\frac{1}{(4\pi)^2}\left[2\hat{\Gamma}^{(1)}_i+\hat{\Gamma}^{(L)}_i(\mu)
\right] \; .
\end{equation} 
The derivation makes use of the relations 
\begin{equation}
\mu\frac{d \hat{\Gamma}^{(L)}_i(\mu)}{d\mu}=-
\frac{\hat{\Gamma}^{(2)}_i}{8\pi^2}
\end{equation} 
that are a mere rewriting of Eqs.~(\ref{eq:WR}). These 115 consistency 
conditions also imply that the chiral double logs can
be calculated from one-loop diagrams alone \cite{Wein79,Col95,BCE1}.

The full result for three flavours can be rewritten in a similar fashion:
\begin{equation}
z^{(3){\rm inf}}_2 = \frac{(c\mu)^{2(d-4)}}{F^2}
\sum_{i=1}^{94} \left(\Gamma^{(2)}_i\Lambda^2 + 
(\Gamma^{(1)}_i+\Gamma^{(L)}_i(\mu,d))\Lambda\right)\int d^d x \,O_i ~.
\end{equation}
The generating functional can be rendered finite by replacing
in
\begin{equation}
{\cal L}_6 = \sum_{i=1}^{94} C_i O_i
\end{equation}
the $C_i$ by
\begin{equation}
\label{replaceCi}
C_i = {(c\mu)^{2(d-4)} \over F^2} \left(
  C_i^r(\mu,d)-\Gamma^{(2)}_i\Lambda^2 
  -(\Gamma^{(1)}_i+\Gamma^{(L)}_i(\mu,d))\Lambda\right)\,.
\end{equation}
The coefficients $\Gamma^{(2)}_i$,
$\Gamma^{(1)}_i$ and $\Gamma^{(L)}_i$
can be found in Table \ref{tab:3f} again in Minkowski conventions.
The structures $O_i$ were derived in Ref. \cite{BCE2} and the correspondence
to the $Y_i$ is listed in Table \ref{tab:3f}.

We present the result for two flavours in the basis of Ref.~\cite{GL84}.
This differs from the reduction of the three-flavour basis of \cite{GL85}
using the Cayley-Hamilton relations for two flavours
by a term proportional to the EOM.
As shown in Eq.~(\ref{liSU2}), this corresponds to a nonzero value
of ${\hat L}_{11}$ in the Lagrangian (\ref{L4N}).
As a consequence, the  contributions of the 1PR
diagrams (c,e,f) of Fig.~\ref{fig:p6diag} are not finite and must
be included. 

The full result for two flavours can be written as
\begin{equation}
z^{(2){\rm inf}}_2 = \frac{(c\mu)^{2(d-4)}}{F^2}\sum_{i=1}^{57} 
\left(\gamma^{(2)}_i\Lambda^2 + 
(\gamma^{(1)}_i+\gamma^{(L)}_i(\mu,d))\Lambda\right)\int d^d x \,P_i
\end{equation}
so that the generating functional is made finite by replacing
in
\begin{equation}
{\cal L}_6 = \sum_{i=1}^{57} c_i P_i
\end{equation}
the $c_i$ by
\begin{equation}
\label{repci}
c_i = {(c\mu)^{2(d-4)} \over F^2} \left(
  c_i^r(\mu,d)-\gamma^{(2)}_i\Lambda^2 
  -(\gamma^{(1)}_i+\gamma^{(L)}_i(\mu,d))\Lambda\right)\,.
\end{equation}
The coefficients $\gamma^{(2)}_i$,
$\gamma^{(1)}_i$ and $\gamma^{(L)}_i$
are given in Table \ref{tab:2f} again in Minkowski conventions.
The structures $P_i$ were derived in Ref. \cite{BCE2} and the correspondence
to the $Y_i$ is listed in Table \ref{tab:2f}.

\setcounter{equation}{0}
\section{Applications}
\label{sec:app}

We now compare our calculation of the infinities with existing
two-loop calculations in CHPT. We also give the expressions
for the counterterm contributions to most of the two-flavour processes
explicitly. Together with formulas in the relevant references this
then provides the full $p^6$ expressions.

In the abnormal-intrinsic-parity sector the infinities and several
$p^6$ quantities have been calculated. But this case involves at most
one-loop contributions and is therefore next-to-leading order.

In the three-flavour case there are only a few calculations available.
The combination of form factors in \cite{PS} was chosen precisely
to have no contributions from the $p^6$ Lagrangian and hence has no infinities
either. All quantities related to vector and axial-vector
two-point functions are also known: the isospin and hypercharge
vector two-point functions in \cite{GK1} and the axial-vector
two-point functions in \cite{GK2}. These results were confirmed in 
\cite{ABT} and in addition the kaonic quantities were calculated.
In all cases the infinities agree with those obtained in this paper.
Explicit expressions for the dependence on the low-energy constants 
$C_i$ can be found in \cite{ABT}.

The first $p^6$ two-loop calculation was $\gamma\gamma\to\pi^0\pi^0$ 
\cite{BGS} followed by $\gamma\gamma\to\pi^+\pi^-$
and the related polarizabilities \cite{Burgi}. In the
latter work also $F_\pi$ and $M_\pi^2$ were obtained. These were later
checked by \cite{BCEGS2,BCEGS1} where also $\pi\pi$ scattering was
evaluated to $O(p^6)$. The vector and scalar form factors of the pion
were evaluated in \cite{BCT} and the form factors of radiative pion decay
in \cite{BT}.

The pion decay constant receives a contribution
\begin{equation}
   r^r_F =   8c^r_{7} + 16c^r_{8} + 8c^r_{9}
\end{equation}
from the Lagrangian of $O(p^6)$ \cite{BCE2}.
The precise definition of $r^r_F$ is given in Eq.~(3.1) of \cite{BCT}.
Similarly, the pion mass has
\begin{equation}
   r^r_M =    - 32c^r_{6} - 16c^r_{7} - 32c^r_{8} - 16c^r_{9} 
 + 48c^r_{10} + 96c^r_{11} 
           + 32c^r_{17} + 64c^r_{18} 
\end{equation}
with $r^r_M$ defined in Eq.~(3.2) of \cite{BCT}.

In $\pi\pi$ scattering there are six additional constants. These
correspond to the contribution from the $p^6$ Lagrangian
to the six possible kinematic coefficients in $A(s,t,u)$. The precise
definition appears in Eq.~(4.14) and App. D of \cite{BCEGS2}. In terms
of the $p^6$ couplings, the six constants are given by
\begin{eqnarray}
  r^r_1 & = &    64c^r_{1} - 64c^r_{2} + 32c^r_{4} - 32c^r_{5} 
+ 32c^r_{6} - 64c^r_{7} - 
         128c^r_{8} - 64c^r_{9} 
\nonumber\\&&
+ 96c^r_{10} + 192c^r_{11} - 64c^r_{14} 
+ 64c^r_{16} + 96c^r_{17}
          + 192c^r_{18} 
\nonumber\\
  r^r_2 & = &    - 96c^r_{1} + 96c^r_{2} + 32c^r_{3} - 32c^r_{4} 
  + 32c^r_{5} - 64
         c^r_{6} + 32c^r_{7} + 64c^r_{8}
\nonumber\\&& 
+ 32c^r_{9} - 32c^r_{13} + 32c^r_{14} - 64c^r_{16} 
\nonumber\\
 r^r_3 &=&    48c^r_{1} - 48c^r_{2} - 40c^r_{3} + 8c^r_{4} 
 - 4c^r_{5} + 8c^r_{6} - 
         8c^r_{12} + 20c^r_{13}
\nonumber\\
 r^r_4 &= &    - 8c^r_{3} + 4c^r_{5} - 8c^r_{6} + 8c^r_{12} - 4c^r_{13} 
\nonumber\\
 r^r_5  & = &    - 8c^r_{1} + 10c^r_{2} + 14c^r_{3} 
\nonumber\\
 r^r_6 & = &   6c^r_{2} + 2c^r_{3} \; \; .
\end{eqnarray}

In the vector form factor of the pion there are two combinations of 
constants:
\begin{eqnarray}
 r^r_{V1} &=&  - 16c^r_{6} - 4c^r_{35} - 8c^r_{53}
\nonumber\\
 r^r_{V2} &=&  - 4c^r_{51} + 4c^r_{53}\; \; .
\end{eqnarray}
Combined with (3.17) and (3.18) of \cite{BCT} this gives the complete
result.
The scalar pion form factor contains the three combinations
\begin{eqnarray}
 r^r_{S1} & = & 3 r^r_M 
\nonumber\\
 r^r_{S2} & = & 32c^r_{6} + 16c^r_{7} + 32c^r_{8} + 16c^r_{9} + 16c^r_{20}
\nonumber\\
 r^r_{S3} & = & - 8c^r_{6}\; \; .
\end{eqnarray}
The first relation is a consequence of the Feynman-Hellman theorem. The
definition and the rest of the full $p^6$ expression can be found in 
Eqs.~(3.6-3.8) and (3.12) of \cite{BCT}.

The decay $\pi(p)\to e\nu\gamma(q)$ is described by two form factors 
$V$ and $A$. Writing the $p^6$ Lagrangian contribution as \cite{BT}
\begin{equation}
A((p-q)^2) = M_\pi^2 r^r_{A1} + (p \cdot q) r^r_{A2}\; \; ,
\end{equation}
we obtain
\begin{eqnarray}
r^r_{A1} & = &  48c^r_{6} - 16c^r_{34} + 8c^r_{35} - 8c^r_{44} + 16
         c^r_{46} - 16c^r_{47} + 8c^r_{50} 
\nonumber\\
r^r_{A2} & = & 8c^r_{44} - 16c^r_{50} + 4 c^r_{51}\; \; .
\end{eqnarray}

In all the above cases the expressions were obtained after the
replacements (\ref{repci}). The infinite parts agree with those
from the explicit calculations using Feynman diagrams\footnote{A small
mistake in \cite{BT} was uncovered in the process of checking
the infinities. An erratum will be published.}.

\setcounter{equation}{0}
\section{Conclusions}
\label{sec:conc}
Chiral perturbation theory, the low-energy effective theory of QCD, is
nowadays being used to calculate processes at the two-loop level
\cite{BCEGS2,PS,GK1,GK2,ABT,BGS,Burgi,BCEGS1,BCT,BT}.
None of those calculations, however, made an attempt at relating the
new low-energy constants to those appearing in other processes, nor could
they check the scale invariance of their final result by using the
(yet unknown) renormalization group equations of the $O(p^6)$ 
low-energy constants. In
Ref. \cite{BCE2} we have given a complete and minimal basis for the
chiral Lagrangian of
order $O(p^6)$, thereby allowing to establish relations between the
constants appearing in different processes. In the present paper we have
completed the formulation of CHPT at the $O(p^6)$ level by calculating the
renormalization group equations for all the new low-energy constants.

We have performed the complete renormalization of the generating functional
of Green functions of quark currents to $O(p^6)$.  We calculated the
divergence structure in chiral $SU(n)$ and then
specialized to the realistic cases $n=2,3$. We have described how to obtain
the divergence structure using heat-kernel techniques, paying special
attention to the influence of terms in the $O(p^4)$ Lagrangian that vanish
by the lowest-order EOM. We have shown how the nonlocal divergences cancel
as is required in any consistent local quantum field theory, by first 
proving it in a general setting and then explicitly verifying it in
our calculation for the case of CHPT.  
The results for $n=2,3$ are obviously of interest for phenomenological
applications, but we stress that the result for generic $n$ has
applications in the case of quenched CHPT \cite{qchpt}, the low-energy
effective theory of QCD in the quenched approximation. As it was shown in
\cite{CP}, the $n$-independent part of the divergences is the result that
one would obtain in a quenched CHPT calculation. It can be used to extract
the two-loop double and single chiral logs appearing in quenched QCD
calculations.

We gave examples of the laborious calculations for three
special cases of one-particle irreducible diagrams
to give an indication of the  algebraic complexity of this calculation.
Our main results, the coefficients of the double- and single-pole
divergences using the standard bases of Ref.~\cite{BCE2} for the chiral
Lagrangians of $O(p^6)$ were presented in the tables in App.~\ref{app:n23}
for $n=3$ and $2$ light flavours and in App.~\ref{app:su(n)} for chiral
$SU(n)$.  

Since several two-loop calculations for specific processes are already
available in the literature, we have compared our results for the
infinities with the published ones and found complete
agreement. Moreover, we have presented the expressions for the
contributions from the $O(p^6)$ Lagrangian to the pion mass and decay
constant, the $\pi \pi$ scattering amplitude, the vector and scalar pion
form factor, and for the $\pi \rightarrow e \nu \gamma$ form factors.

\vfill
\section*{Acknowledgements}

We thank J\"urg Gasser for many helpful discussions.

\setcounter{equation}{0}
\newcounter{zahler}
\addtocounter{zahler}{1}
\renewcommand{\thesection}{\Alph{zahler}}
\renewcommand{\theequation}{\Alph{zahler}.\arabic{equation}}

\appendix

\section{Singular parts of the products of propagators} 
\label{app:JO}
The short-distance singularities of the following products of
propagator functions $G_0$, $R_1$, $R_2$ (defined in (\ref{propfunc}))
and their derivatives are needed for calculating the divergences of
the generating functional $z_2$ in (\ref{z2}). We have checked the
results of Ref.~\cite{JO} and reproduce them here for convenience
($\epsilon = 4 - d$).
 
\bea
G_0(x)^2 &\sim& \frac{1}{16 \pi^2} {2 \over \epsilon} \delta^d(x)  \\
G_0(x) \partial_\mu G_0(x) &\sim& \frac{1}{16 \pi^2} {1 \over \epsilon}
\partial_\mu \delta^d(x)  \\
\partial_\mu G_0(x) \partial_\nu G_0(x) &\sim& \frac{1}{16 \pi^2} {1 \over
 6 \epsilon} \left( \delta_{\mu \nu} \partial^2+2 \partial_\mu
  \partial_\nu \right) \delta^d(x)  \\
\partial_\mu G_0(x) \partial_\nu R_1(x) &\sim& \frac{1}{16 \pi^2} {1 \over
 2 \epsilon} \delta_{\mu \nu} \delta^d(x)  \\
G_0(x)^3 &\sim& \frac{1}{(16 \pi^2)^2} {1 \over 2\epsilon}
\partial^2 \delta^d(x)  \\
G_0(x)^2 R_1(x) &\sim& \frac{1}{(16 \pi^2)^2} \left( {2 \over \epsilon^2}+
  {1 \over \epsilon} \right)  \delta^d(x)  \\
G_0(x)^2\partial_\mu G_0(x) &\sim& \frac{1}{(16 \pi^2)^2} {1 \over
  6\epsilon} \partial_\mu \partial^2 \delta^d(x)  \\
G_0(x) \partial_\mu G_0(x) R_1(x) &\sim& \frac{1}{(16 \pi^2)^2} \left( {1
    \over \epsilon^2}+ {1 \over 4 \epsilon} \right) \partial_\mu
\delta^d(x)  \\ 
G_0(x)^2 \partial_\mu R_1(x) &\sim& \frac{1}{(16 \pi^2)^2} {1 \over
  2\epsilon} \partial_\mu \delta^d(x)  \\
\partial_\mu G_0(x) \partial_\nu G_0(x) G_0(x) &\sim& \frac{1}{(16
  \pi^2)^2} {1 \over 96\epsilon} \left(\delta_{\mu
    \nu}\partial^2 +4  \partial_\mu \partial_\nu \right)\partial^2
\delta^d(x)  \\ 
\partial_\mu G_0(x) \partial_\nu G_0(x) R_1(x) &\sim& \frac{1}{(16
  \pi^2)^2} \left( {1 \over 6\epsilon^2} - {1 \over 72
    \epsilon} \right) \left( \delta_{\mu \nu} \partial^2 +2 \partial_\mu
  \partial_\nu \right) \delta^d(x)  \nn 
    && \\ 
G_0(x) \partial_\mu G_0(x) \partial_\nu R_1(x) &\sim& \frac{1}{(16
  \pi^2)^2}  {1 \over 12\epsilon} \left( \delta_{\mu
    \nu}\partial^2  +2 \partial_\mu \partial_\nu \right) \delta^d(x) \\  
G_0(x) \partial_\mu R_1(x) \partial_\nu R_1(x) &\sim& \frac{1}{(16
  \pi^2)^2} {1 \over 4\epsilon} \delta_{\mu \nu} \delta^d(x)  \\  
\partial_\mu G_0(x) \partial_\nu R_1(x) R_1(x) &\sim& \frac{1}{(16
  \pi^2)^2} \left( {1 \over 2\epsilon^2} + {1 \over 8 \epsilon}
  \right) \delta_{\mu \nu} \delta^d(x)  \\ 
\partial_\mu G_0(x) \partial_\nu G_0(x) R_2(x) &\sim& -\frac{1}{(16
  \pi^2)^2} \left( {1 \over 2\epsilon^2} + {1 \over 8
    \epsilon} \right) \delta_{\mu \nu} \delta^d(x)  \\ 
G_0(x) \partial_\mu G_0(x) \partial_\nu R_2(x) &\sim& \frac{1}{(16
  \pi^2)^2} \left( {1 \over 2\epsilon^2} + {1 \over 8
    \epsilon} \right) \delta_{\mu \nu} \delta^d(x)  \\ 
\partial_\mu \partial_\nu G_0(x) G_0(x) &\sim& {1 \over 16 \pi^2} {1 \over
  6\epsilon} \left( 4\partial_\mu \partial_\nu - 
  \delta_{\mu \nu} \partial^2 \right) \delta^d(x)  \\
\partial_\mu \partial_\nu G_0(x) R_1(x) &\sim& -{1 \over 16 \pi^2} {1 \over
  2\epsilon} \delta_{\mu \nu} \delta^d(x)  \\
G_0(x) \partial_\mu \partial_\nu R_1(x) &\sim& -{1 \over 16 \pi^2} {1 \over
  2\epsilon} \delta_{\mu \nu} \delta^d(x)  \\
\partial_\mu \partial_\nu G_0(x) G_0(x)^2 &\sim& {1 \over (16 \pi^2)^2} {1
  \over 48\epsilon} \left( 4\partial_\mu \partial_\nu - 
  \delta_{\mu \nu} \partial^2 \right) \partial^2 \delta^d(x)  \\
\partial_\mu \partial_\nu G_0(x) G_0(x) R_1(x) &\sim& {1 \over (16
  \pi^2)^2} \left( {2 \over 3\epsilon^2}+{1 \over 9\epsilon}
\right) \left( \partial_\mu \partial_\nu - {1 \over d}
   \delta_{\mu \nu} \partial^2 \right) \delta^d(x) ~~~~~~  \\
\partial_\mu \partial_\nu G_0(x) R_1(x)^2 &\sim&- {1 \over (16 \pi^2)^2}
\left( {1 \over \epsilon^2} + {1 \over 4\epsilon} \right)
\delta_{\mu \nu} \delta^d(x)  \\ 
G_0(x)^2 \partial_\mu \partial_\nu R_1(x) &\sim& {1 \over (16 \pi^2)^2}
{1 \over 6\epsilon} \left( \partial_\mu \partial_\nu -
  \delta_{\mu \nu} \partial^2 \right)  \delta^d(x)  \\ 
G_0(x) \partial_\mu \partial_\nu R_1(x) R_1(x) &\sim&-{1 \over (16
  \pi^2)^2} \left( {1 \over 2\epsilon^2} + {3 \over 8
    \epsilon} \right) \delta_{\mu \nu} \delta^d(x)  \\ 
\partial_\mu \partial_\nu G_0(x) G_0(x) R_2(x) &\sim&0  \\
G_0(x)^2 \partial_\mu \partial_\nu R_2(x) &\sim& -{1 \over (16 \pi^2)^2}
\left( {1 \over \epsilon^2} + { 1 \over 4 \epsilon} \right)
\delta_{\mu \nu} \delta^d(x) \; \; .
\label{singprod}
\eea

\setcounter{equation}{0}
\addtocounter{zahler}{1}

\section{Cancellation of subdivergences}
\label{app:subdivs}

On the basis of the general proof given in Secs.~\ref{sec:subdivs} and 
\ref{sec:Wein}, all nonlocal subdivergences 
appearing at intermediate stages of the calculation have to cancel in the 
final result. In principle, one could drop them whenever they are 
produced during the calculation. We have chosen to keep them and explicitly 
check that they cancel in the final result.
To arrive at the complete cancellation, one has to go through a series 
of nontrivial manipulations that we are now going to describe briefly.

It should be clear from our general discussion in
Secs.~\ref{sec:subdivs} and \ref{sec:Wein}
that all nonlocal subdivergences appear in this
framework as single poles in $d-4$ proportional to 
$\Gb$ and derivatives thereof. Since the manipulations we are going to
describe produce terms with less derivatives than the term one starts from,
we proceed in the discussion from the terms with the highest number of 
derivatives (two) down to the terms without derivatives. 
In what follows we will use the symmetry property of the
matrix $\Gb$:
\be
\label{eq:Gbsym}
\Gb_{ab}(x,y) = \Gb_{ba}(y,x)~.
\ee
We will also use the abbreviation $\mbox{tr}(M(x,y))=
\int d^dx  d^dy\, \delta^d(x-y) M(x,y)$.
 
\subsection{Terms with two derivatives acting on $\Gb$}
First we consider those terms where the indices of the two derivatives are 
contracted. By partial integration in the generating functional, one
can bring all such terms into the form $d^2 \Gb$. We can then use the 
identity \cite{JO}
    \be
    \triangle_x \Gb(x,y)_{|_{y=x}} = {1 \over 16 \pi^2} a_2(x,x) ~,
    \ee
which for this purpose we use as
\be\label{d2Gb}
\lim_{y\to x} d_{x}^2 \Gb(x,y) = \sigma(x)\Gb(x,x)-{1 \over 16 \pi^2} a_2(x,x) ~.
\ee
So the terms containing $d^2 \Gb$ produce both a local ($a_2$) and a
nonlocal ($\sigma \Gb$) divergence.

    Next we consider terms where the indices of the two derivatives are not 
    contracted. There are three such terms:
    \[ d_\mu^x d_\nu^x \Gb(x,y), \; d_\mu^x d_\nu^y \Gb(x,y) , \; 
   d_\mu^y d_\nu^y  \Gb(x,y) ~. \]

    By partial integration we bring all these terms into
    $d_\mu^x d_\nu^y \Gb(x,y)$.
    This term is then manipulated in the following way:
    \bea
\lefteqn{
    \mbox{tr} \left(d_\mu^x d_\nu^y \Gb_{ab}(x,y) \; A_{ba\,\mu \nu}(x,y)\right) 
\;=}&&\nonumber\\ &&
{1 \over 2} \mbox{tr} 
    \left[ \left( d_\mu^x d_\nu^y \Gb_{ab}(x,y)  
   + d^x_\nu d^y_\mu \Gb_{ab}(x,y) 
\right) \; A_{ba\,\mu \nu}(x,y) \right] 
    \nonumber \\
    &+&  {1 \over 2} \mbox{tr} \left(\gamma_{ca\,\mu
        \nu}(x) \Gb_{ab}(x,y) A_{bc\,\mu \nu}(x,y)  \right)  
    \nonumber \\
    &-&  {1 \over 2} \mbox{tr} \left[ (d^x_\mu \Gb_{ab}(x,y)) 
(\hat d_\nu A_{ba\,\mu \nu}(x,y) ) -
      (d^x_\nu \Gb_{ab}(x,y)) 
      ( \hat d_\mu A_{ba\,\mu \nu}(x,y)) \right] ~,
    \eea
using
\begin{eqnarray} 
[d^x_\mu,d^x_\nu] \Gb_{ab}(x,y) &=& \gamma_{ac\,\mu\nu}(x) \Gb_{cb}(x,y) \nn
\gamma_{ab\,\mu\nu}&=& - \frac{1}{2}\lgl 
\Gamma_{\mu\nu}[\lambda_a,\lambda_b] \rgl \nn
\Gamma_{\mu\nu} &=& \partial_\mu \Gamma_\nu - \partial_\nu \Gamma_\mu 
+[\Gamma_\mu,\Gamma_\nu]~.
\end{eqnarray} 
Moreover, we have defined the total derivative
\be
\hat d_\mu X(x,y) = d^x_\mu X(x,y) + d^y_\mu X(x,y)~.
\ee
    The aim of this manipulation is to separate the part of the
    two-derivative terms that is symmetric with respect to exchanges of
    the two ``flavour'' indices $a,b$ (running from $1$ to $n^2 -
    1$) from the one
    that is antisymmetric. The antisymmetric part produces the
    $\gamma_{\mu \nu}$ term. Then we have a term that is produced by 
    partial integration where only one derivative is left on $\Gb$.
    These terms have to be manipulated separately, as we are now going to
    describe. 

\subsection{Terms with one derivative on $\Gb$}
    Also in this case the derivative can act on $\Gb$ from two different
    sides: $d^x_\mu \Gb(x,y)$ and $d^y_\mu\Gb(x,y)$. Again we separate 
symmetric and antisymmetric parts:
    \be
    d^x_\mu \Gb(x,y) = {1 \over 2} \left[ d^x_\mu \Gb(x,y) 
     + d^y_\mu\Gb(x,y)
    \right] + {1 \over 2} \left[ d^x_\mu \Gb(x,y) - d^y_\mu\Gb(x,y)\right] \;,
    \ee
    and analogously for $d^y_\mu\Gb(x,y)$. Then we partially integrate the
    symmetric term:
    \be
    \mbox{tr} \left[ \left( d^x_\mu \Gb_{ab}(x,y) + d^y_\mu\Gb_{ab}(x,y)\right)
  B_{ba\,\mu}(x,y)
    \right] = -  
    \mbox{tr} \left[ \Gb_{ab}(x,y) 
  \left( \hat d_\mu B_{ba\,\mu}(x,y) \right) \right] \; ,
    \ee
    to give a term with no derivatives on $\Gb$ that we are going to
    analyse next.

\subsection{Terms without derivatives on $\Gb$}
On such terms no manipulation is necessary because of the symmetry
property of $\Gb$, Eq.~(\ref{eq:Gbsym}), which implies
that in the coincidence limit $y \to x$ $\Gb$ is a symmetric matrix.

\subsection{Summary of the procedure}
To summarize, the treatment of the divergent terms containing the nonlocal
part of the propagator $\Gb$ requires a few simple manipulations to bring
everything into the following form:
\bea
\mbox{tr} \left[ \left( d^x_\mu d^y_\nu \Gb_{ab}(x,y)  + d^x_\nu d^y_\mu
\Gb_{ab}(x,y) 
 \right) A_{ba\,\mu\nu}(x,y) \right] +\nonumber\\ \mbox{tr} \left[ 
\left(d^x_\mu \Gb_{ab}(x,y) -  d^y_\mu \Gb_{ab}(x,y)  \right)
B_{ba\,\mu}(x,y) \right] + \mbox{tr} \left( \Gb_{ab}(x,y) C_{ba}(x,y) \right)
 \; .
\eea
If in the coincidence limit $A_{\mu \nu}$ and $C$ are antisymmetric 
and $B_\mu$ is symmetric in the ``flavour'' indices, the whole
expression vanishes. We have explicitly checked that this is what happens.

\setcounter{equation}{0}
\addtocounter{zahler}{1}
\section{Divergences for $n=3$ and $2$}
\label{app:n23}

\renewcommand{\arraystretch}{1.1}
\setlength{\LTcapwidth}{\textwidth}

\newcommand{\extraline}{$\\ & & &$}
\begin{longtable}[c]{|c|c|c|l|}
\hline
$O_i(Y_i)$ & $\Gamma^{(2)}_i$ & $16\pi^2\Gamma^{(1)}_i$ & 
\hspace*{3cm} $\Gamma^{(L)}_i$\\
\hline
\hline
\endhead
\hline
\caption[]{\rule{0cm}{2em}}
\endfoot
\hline
\caption[Infinities]{\label{tab:3f} 
\rule{0cm}{2em} The coefficients $\Gamma^{(2)}_i$ of the double-pole
and $\Gamma^{(1)}_i$, $\Gamma^{(L)}_i$ of the single-pole
divergences for the three-flavour case in Minkowski space.
The number in the first column indicates the $O_i$ with
the corresponding $Y_i$ of Table \ref{tab:nf1} in brackets.
$O_{94}=\det\chi+\det\chi^\dagger$.}
\endlastfoot
1(1) & $ 17/32 $ & $ 71/768$ &
 $ 7/3\,L^r_1 + 35/6\,L^r_2 + 9/4\,L^r_3 - L^r_9 $ \\

2(2) & $ 15/64 $ & $ 1/128$ & $ 3\,L^r_1 + L^r_2 + 5/6\,L^r_3 $ \\

3(3) & $  - 1/4 $ & $  - 43/768$ &
 $ - L^r_1 - 5/2\,L^r_2 - 1/2\,L^r_3 + 1/4\,L^r_9  $ \\

4(5) & $ 9/32 $ & $ 79/768$ & $ 4/3\,L^r_1 + 10/3\,L^r_2 + 3/4\,L^r_3 - 3/4
  L^r_9 $ \\

5(7) & $  - 7/8 

$ & $  - 85/384$ & $ - 8/3\,L^r_1 - 19/3\,L^r_2 - 59/36\,L^r_3 + 
  2\,L^r_4 \extraline- 3/2\,L^r_5 $ \\

6(8) & $  - 2/3 

$ & $ 89/576$ & $ - 10/3\,L^r_1 - 16/9\,L^r_2 - 65/54\,L^r_3 - 4
  L^r_4 \extraline- 1/2\,L^r_5 $ \\

7(9) & $ 9/16 

$ & $ 9/32$ & $ 6\,L^r_1 + 6\,L^r_2 + 3\,L^r_3 - 3\,L^r_4 - 1/2
  L^r_5 $ \\

8(11) & $  - 21/32 

$ & $ 1/12$ & $ - 2/3\,L^r_1 - 13/3\,L^r_2 - 23/36\,L^r_3 + L^r_4
  \extraline - 3/2\,L^r_5 $ \\

9(12) & $ 1/2 

$ & $ 3/64$ & $ 2\,L^r_1 + 5\,L^r_2 - L^r_4 $ \\

10(13) & $  - 1/4 

$ & $  - 41/192$ & $ - 8/3\,L^r_1 - 13/3\,L^r_2 + 1/36\,L^r_3 + 
  3/2\,L^r_5 $ \\

11(14) & $ 29/48 

$ & $ 19/576$ & $ 16/3\,L^r_1 + 7/9\,L^r_2 + 157/108\,L^r_3 + 
  3\,L^r_4 \extraline+ 1/2\,L^r_5 $ \\

12(17) & $ 19/64 

$ & $  - 13/768$ & $ 2/3\,L^r_1 + 4/3\,L^r_2 + 8/9\,L^r_3 + 3/
  4\,L^r_5 $ \\

13(18) & $ 67/192 

$ & $  - 1/2304$ & $ 8/3L^r_1 + 8/9\,L^r_2 + 23/27\,L^r_3 + 
  3/2\,L^r_4 \extraline + 1/4\,L^r_5 $ \\

14(19) & $  - 29/288 

$ & $  - 479/4608$ & $ - 79/18\,L^r_1 + 1/36\,L^r_2 - 37/72
  L^r_3 
\extraline + 4/9\,L^r_4 + 4\,L^r_6 - 2\,L^r_7 - 3/2\,L^r_8 $ \\

15(20) & $ 5/72 

$ & $ 31/192$ & $ 44/9\,L^r_1 + 11/9\,L^r_2 + 53/36\,L^r_3  
\extraline+ 
  19/18\,L^r_4 - 1/9\,L^r_5 - 7\,L^r_6 - L^r_8 $ \\

16(21) & $ 73/576 

$ & $ 43/768$ & $ 26/9\,L^r_1 + 13/18\,L^r_2 + 61/72\,L^r_3  
\extraline- 
  17/18\,L^r_4 + 1/4\,L^r_5 - L^r_6 - 1/2\,L^r_8 $ \\

17(23) & $ 25/576 

$ & $ 55/4608$ & $ - 35/18\,L^r_1 + 23/36\,L^r_2 + 1/18\,L^r_3
    
\extraline+ 2/9\,L^r_4 - 1/12\,L^r_5 + 2\,L^r_6 + 2\,L^r_7 $ \\

18(24) & $  - 31/576 

$ & $ 235/10368$ & $ 19/9\,L^r_1 - 2/9\,L^r_2 + 1/36\,L^r_3 - 
  2/9\,L^r_4  
\extraline- 1/2\,L^r_5 - 2\,L^r_6 + L^r_8 $ \\

19(25) & $ 11/1944 

$ & $ 1517/93312$ & $ 44/81\,L^r_1 + 8/81\,L^r_2 + 10/81
  L^r_3 + 1/3\,L^r_4  
\extraline- 1/4\,L^r_5 + 4/27\,L^r_6 - 32/27\,L^r_7 - 1/6\,L^r_8 $ \\

20(26) & $ 13/1296 

$ & $  - 1517/62208$ & $ - 22/27\,L^r_1 - 4/27\,L^r_2 - 5/27
  L^r_3\extraline + 1/3\,L^r_4  
+13/12\,L^r_5 -17/9\,L^r_6\extraline +1/9\,L^r_7 -31/18\,L^r_8 $ \\

21(27) & $ 59/3888 

$ & $ 1517/186624$ & $ 22/81\,L^r_1 + 4/81\,L^r_2 + 5/81
  L^r_3 + 7/9\,L^r_4  
\extraline- 1/27\,L^r_5 - 31/27\,L^r_6 + 5/27\,L^r_7 $ \\

22(28) & $  - 89/192 

$ & $  - 5/144$ & $ - 16/3\,L^r_1 - 7/3\,L^r_2 - 2\,L^r_3 - 2/3
  L^r_4\extraline + 1/4\,L^r_5 $ \\

23(29) & $ 3/16 

$ & $ 15/256$ & $ 3\,L^r_1 + 3/2\,L^r_2 + L^r_3 - 1/2\,L^r_5 $ \\

24(30) & $ 27/32 

$ & $ 49/288$ & $ 62/9\,L^r_1 + 47/9\,L^r_2 + 3\,L^r_3 + 1/2
  L^r_5 \extraline - 1/2\,L^r_9 $ \\

25(31) & $  - 25/24 

$ & $  - 173/576$ & $ - 10\,L^r_1 - 11\,L^r_2 - 5\,L^r_3 + 4/3
  L^r_4  
\extraline+ L^r_5 + 3/2\,L^r_9 $ \\

26(33) & $ 475/576 

$ & $ 1397/6912$ & $ 191/27\,L^r_1 + 379/54\,L^r_2 + 15/4
  L^r_3 \extraline- 4/3\,L^r_4  
+ 5/6\,L^r_5 - 2\,L^r_6 + 4\,L^r_7\extraline - 3/2\,L^r_8 - 2/3\,L^r_9 $ \\

27(34) & $  - 67/96 

$ & $  - 67/648$ & $ - 134/27\,L^r_1 - 83/27\,L^r_2 - 19/9
  L^r_3 \extraline+ 2/3\,L^r_4  
- L^r_5 - 7\,L^r_7 - L^r_8 + 1/6\,L^r_9 $ \\

28(35) & $ 31/192 

$ & $ 1/6912$ & $ 58/27\,L^r_1 + 28/27\,L^r_2 + 3/4\,L^r_3 - 
  L^r_7  
\extraline- 1/2\,L^r_8 - 1/12\,L^r_9 $ \\

29(37) & $ 91/288 

$ & $  - 59/1728$ & $ 37/27\,L^r_1 - 49/54\,L^r_2 + 1/3\,L^r_3
   - 2/3\,L^r_4  
\extraline+ 4/3\,L^r_5 + 2\,L^r_6 + 2\,L^r_7 + 5/12\,L^r_9 $ \\

30(38) & $ 85/96 

$ & $ 91/3456$ & $ 98/27\,L^r_1 + 74/27\,L^r_2 + 16/9\,L^r_3
   +\extraline 17/3\,L^r_4  
+ 7/18\,L^r_5 - 2\,L^r_7 + L^r_8 - 1/6\,L^r_9 $ \\

31(39) & $ 1025/2592 

$ & $ 2359/62208$ & $ 62/27\,L^r_1 + 44/27\,L^r_2 + 34/27
  L^r_3 + L^r_4  
\extraline+ 3/4\,L^r_5 - 14/9\,L^r_6 + 22/9\,L^r_7 - 1/6\,L^r_8 $ \\

32(40) & $ 569/2592 

$ & $  - 851/62208$ & $ 50/27\,L^r_1 + 20/27\,L^r_2 + 2/3
  L^r_3 + L^r_4  
\extraline+ 1/4\,L^r_5 - 2/9\,L^r_6 - 11/9\,L^r_7 - 13/18\,L^r_8 $ \\

33(41) & $  - 31/81 

$ & $  - 451/15552$ & $ - 56/27\,L^r_1 - 32/27\,L^r_2 - 26/27
  \,L^r_3 - L^r_4  
\extraline- 1/2\,L^r_5 + 8/9\,L^r_6 - 37/9\,L^r_7 - L^r_8 $ \\

34(43) & $  - 13/32 

$ & $  - 31/2304$ & $ - L^r_1 - 3/2\,L^r_2 - 11/12\,L^r_3 + L^r_4
   - 3/2\,L^r_5 $ \\

35(44) & $ 13/48 

$ & $ 43/10368$ & $ 2/3\,L^r_1 + L^r_2 + 11/18\,L^r_3 - 2/3
  L^r_4 + L^r_5 $ \\

36(45) & $  - 29/24 

$ & $  - 1/12$ & $ - 16/3\,L^r_1 - 8/3\,L^r_2 - 35/18\,L^r_3 \extraline - 
  25/3\,L^r_4- L^r_5 $ \\

37(46) & $

$ & $ 173/20736$ & $ $ \\

38(47) & $  - 5/32 

$ & $ 49/2304$ & $ - 5/6\,L^r_5 $ \\

39(48) & $  - 9/32 $ & $ $ & $  - 8/3\,L^r_4 - 11/18\,L^r_5 $ \\

40(49) & $  - 15/16 

$ & $  - 379/1152$ & $ - 2/3\,L^r_1 - 11\,L^r_2 - 31/12\,L^r_3
   + L^r_9 $ \\

41(50) & $  - 3/64 

$ & $ 71/256$ & $ - 6\,L^r_1 + 5/2\,L^r_2 - 2/3\,L^r_3 $ \\

42(52) & $  - 3/4 

$ & $  - 233/1152$ & $ - 26/3\,L^r_2 - 11/8\,L^r_3 + 1/2\,L^r_9
   $ \\

43(53) & $ 27/32 

$ & $ 27/128$ & $ 9\,L^r_2 $ \\

44(54) & $  - 3/8 

$ & $ 53/1152$ & $ - 2/3\,L^r_1 - L^r_2 + 5/4\,L^r_3 - 2\,L^r_9 $ \\

45(57) & $ 87/128 

$ & $ 47/512$ & $ 6\,L^r_1 + 17/4\,L^r_2 + 13/6\,L^r_3 $ \\

46(58) & $ 3/16 

$ & $ 281/1152$ & $ 2/3\,L^r_1 + 7/3\,L^r_2 - 7/24\,L^r_3 - 1/
  2\,L^r_9 $ \\

47(60) & $  - 21/32 

$ & $  - 451/1152$ & $ 2/3\,L^r_1 - 26/3\,L^r_2 - 3/8\,L^r_3
   + L^r_9 $ \\

48(64) & $  - 1/64 

$ & $  - 65/192$ & $ - 1/3\,L^r_1 - 11/6\,L^r_2 + 11/12\,L^r_3
\extraline   + 11/8\,L^r_9 $ \\

49(65) & $  - 39/32 

$ & $ 5/128$ & $ - 12\,L^r_1 - 7\,L^r_2 - 13/3\,L^r_3 $ \\

50(66) & $  - 13/24 

$ & $  - 107/288$ & $ - 2/3\,L^r_1 - 5\,L^r_2 + 4/3\,L^r_3 - 2/3
  \,L^r_4  
\extraline- 1/2\,L^r_5 + 3/4\,L^r_9 $ \\

51(67) & $  - 55/96 

$ & $  - 43/144$ & $ - 4/3\,L^r_1 - 26/3\,L^r_2 - 17/6\,L^r_3 - 
  2/3\,L^r_4  
\extraline- 1/2\,L^r_5 + 7/2\,L^r_9 $ \\

52(68) & $ 41/48 

$ & $ 197/1152$ & $ - 4/3\,L^r_1 + 8\,L^r_2 + 5/6\,L^r_3 + 2/3
  L^r_4  
\extraline+ 1/2\,L^r_5 + 1/4\,L^r_9 $ \\

53(71) & $ 7/48 

$ & $  - 7/576$ & $ - 2/3\,L^r_1 - 1/6\,L^r_3 + 1/3\,L^r_4 + 1/4
  \,L^r_5  
\extraline- 7/8\,L^r_9 - 7/4\,L^r_{10} $ \\

54(72) & $  - 3/64 

$ & $  - 7/256$ & $ - 1/2\,L^r_2 - 1/6\,L^r_3 $ \\

55(73) & $  - 7/48 

$ & $ 43/576$ & $ 2/3\,L^r_1 + 5/12\,L^r_3 - 1/3\,L^r_4 - 1/4
  L^r_5  
\extraline- 11/8\,L^r_9 - 1/2\,L^r_{10} $ \\

56(75) & $ 1/24 

$ & $ 1/6$ & $ 2/3\,L^r_2 - 2/3\,L^r_3 + 2/3\,L^r_4  
\extraline+ 1/2\,L^r_5
   - 5/4\,L^r_9 $ \\

57(76) & $  - 1/3 

$ & $ 7/32$ & $ 2\,L^r_1 - 1/3\,L^r_2 + 5/6\,L^r_3 - 4/3\,L^r_4
   - L^r_5\extraline - L^r_9 $ \\

58(77) & $ 3/16 

$ & $  - 11/64$ & $ 2\,L^r_2 + 2/3\,L^r_3 $ \\

59(78) & $ 7/48 

$ & $ 5/48$ & $ - L^r_1 - 1/6\,L^r_2 - 7/12\,L^r_3 + 1/3\,L^r_4  
\extraline+ 
  1/4\,L^r_5 + 9/8\,L^r_9 $ \\

60(80) & $
$ & $  - 3/32$ & $ $ \\

61(81) & $ $ & $ $ & $  - 3/4\,L^r_9 - 3/4\,L^r_{10} $ \\

62(82) & $ $ & $ $ & $  - 1/4\,L^r_9 - 1/4\,L^r_{10} $ \\

63(83) & $  - 1/4 

$ & $ 5/384$ & $ 1/3\,L^r_1 - 11/6\,L^r_2 - 5/36\,L^r_3 - 3/4
  L^r_5 \extraline + 3/8\,L^r_9 $ \\

64(84) & $  - 4/3 

$ & $  - 5/144$ & $ - 32/3\,L^r_1 - 32/9\,L^r_2 - 157/54\,L^r_3
   - 6\,L^r_4  
\extraline- L^r_5 + 1/2\,L^r_9 $ \\

65(85) & $  - 1/2 

$ & $ 5/192$ & $ 2/3\,L^r_1 - 11/3\,L^r_2 - 16/9\,L^r_3 - 3/2
  L^r_5 \extraline + 3/4\,L^r_9 $ \\

66(86) & $  - 1/32 

$ & $ 115/1152$ & $ 2/3\,L^r_1 + L^r_2 - 5/6\,L^r_3 - 5/4\,L^r_9
   $ \\

68(88) & $  - 3/4 

$ & $  - 7/64$ & $ - 12\,L^r_1 - 2\,L^r_2 - 8/3\,L^r_3 $ \\

69(89) & $  - 1/4 

$ & $ 29/1152$ & $ - 2/3\,L^r_1 - L^r_2 - 5/3\,L^r_3 - L^r_9 $ \\

70(90) & $  - 35/48 

$ & $  - 389/2304$ & $ - 8/3\,L^r_1 - 17/3\,L^r_2 - 5/2\,L^r_3
   + 1/3\,L^r_4  
\extraline+ 1/4\,L^r_5 + 5/8\,L^r_9 + 7/4\,L^r_{10} $ \\

71(91) & $  - 15/64 

$ & $  - 7/128$ & $ - 3\,L^r_1 - L^r_2 - 5/6\,L^r_3 $ \\

72(92) & $ 1/6 

$ & $ 251/2304$ & $ 4/3\,L^r_1 + 7/3\,L^r_2 + 3/4\,L^r_3 - 1/3
  \,L^r_4  
\extraline- 1/4\,L^r_5 + 1/8\,L^r_9 + 1/2\,L^r_{10} $ \\

73(94) & $ 31/48 

$ & $ 385/1152$ & $ 10/3\,L^r_1 + 9\,L^r_2 + 11/6\,L^r_3 - 2/3
  \,L^r_4  
\extraline- 1/2\,L^r_5 - 7/4\,L^r_9 $ \\

74(95) & $  - 7/8 

$ & $  - 13/128$ & $ - 8/3\,L^r_1 - 20/3\,L^r_2 - 4\,L^r_3 - L^r_9
   $ \\
75(96) & & & \\
76(97) & $  - 43/96 

$ & $  - 217/2304$ & $ - 5/3\,L^r_1 - 9/2\,L^r_2 - 13/6\,L^r_3
   + 1/3\,L^r_4  
\extraline+ 1/4\,L^r_5 - 1/8\,L^r_9 $ \\

77(99) & $  - 3/4 

$ & $  - 7/64$ & $ - 12\,L^r_1 - 2\,L^r_2 - 8/3\,L^r_3 $ \\

78(100) & $  - 5/24 

$ & $ 49/1152$ & $ 1/3\,L^r_1 - 1/6\,L^r_2 + 1/4\,L^r_3 - 1/3
  L^r_4  
\extraline- 1/4\,L^r_5 - 9/8\,L^r_9 $ \\

79(101) & $ 1/48 

$ & $  - 11/576$ & $ - 1/3\,L^r_1 + 1/6\,L^r_2 - 1/4\,L^r_3 + 1/
  3\,L^r_4  
\extraline+ 1/4\,L^r_5 - 3/8\,L^r_9 $ \\

80(102) & $  - 25/64 

$ & $  - 17/768$ & $ - 2/3\,L^r_1 - 4/3\,L^r_2 - 8/9\,L^r_3  
\extraline- 3/
  4\,L^r_5 + 3/4\,L^r_{10} $ \\

81(103) & $  - 73/192 

$ & $  - 101/2304$ & $ - 8/3\,L^r_1 - 8/9\,L^r_2 - 23/27\,L^r_3
   - 3/2\,L^r_4  
\extraline- 1/4\,L^r_5 + 1/4\,L^r_{10} $ \\

82(104) & $ 7/96 

$ & $ 5/144$ & $ - 1/6\,L^r_1 + 1/12\,L^r_2 - 1/8\,L^r_3 + 1/6
  L^r_4 
\extraline+ 1/8\,L^r_5 + 5/16\,L^r_9 $ \\

83(105) & $ 115/192 

$ & $ 359/2304$ & $ 3\,L^r_1 + 11/2\,L^r_2 + 97/36\,L^r_3 - 5/
  3\,L^r_4 \extraline + 1/4\,L^r_5 $ \\

84(106) & $ 5/3 

$ & $ 215/1152$ & $ 50/3\,L^r_1 + 43/9\,L^r_2 + 133/27\,L^r_3
   + 6\,L^r_4 \extraline+ 1/2\,L^r_9 $ \\

85(107) & $ 5/8 

$ & $  - 1/192$ & $ 2/3\,L^r_1 + 7/3\,L^r_2 + 31/18\,L^r_3 + 3/
  2\,L^r_5 \extraline+ 3/4\,L^r_9 $ \\

86(108) & $ 31/24 

$ & $ 5/72$ & $ 32/3\,L^r_1 + 16/9\,L^r_2 + 79/27\,L^r_3 + 6
  L^r_4  
\extraline+ L^r_5 + 1/2\,L^r_9 $ \\

87(109) & $  - 1/16 

$ & $ 3/128$ & $ - 1/2\,L^r_9 $ \\

88(110) & $ 1/6 

$ & $  - 31/576$ & $ - 2/3\,L^r_1 + 1/3\,L^r_2 - 1/2\,L^r_3 + 2/
  3\,L^r_4  
\extraline+ 1/2\,L^r_5 + 1/4\,L^r_9 $ \\

89(111) & $  - 7/32 

$ & $ 1/32$ & $ 2/3\,L^r_1 - 1/3\,L^r_2 + 1/2\,L^r_3 - 7/4\,L^r_9
   $ \\

90(112) & $ 13/96 

$ & $  - 37/576$ & $ 2/3\,L^r_4 + 1/2\,L^r_5 $ \\

91(113) & $ 
$ & $  - 49/576$ & $ $ \\

92(114) & $  - 1/2 

$ & $  $ & $- 4\,L^r_9 $\\

93(115) & $ 1/8 $ & $ $ & $ L^r_9 $\\

94 & $  - 119/162 $  & $  - 1517/7776$ & $ - 176/27\,L^r_1 - 32/27\,L^r_2 - 40/
  27\,L^r_3 \extraline - 4\,L^r_4  
- 16/9\,L^r_6 - 88/9\,L^r_7 $ \\
\end{longtable}

\renewcommand{\extraline}{$\\ & & &$}
\begin{longtable}[c]{|c|c|c|l|}
\hline
$P_i(Y_i)$ & $\gamma^{(2)}_i$ & $16\pi^2\gamma^{(1)}_i$ & 
\hspace*{3cm} $\gamma^{(L)}_i$\\
\hline
\hline
\endhead
\hline
\caption[]{\rule{0cm}{2em}}
\endfoot
\hline
\caption[Infinities]{\label{tab:2f} 
\rule{0cm}{2em} The coefficients $\gamma^{(2)}_i$ of the double-pole
and $\gamma^{(1)}_i$, $\gamma^{(L)}_i$ of the single-pole
divergences for the two-flavour case in Minkowski space.
The first number in the first column refers to the $P_i$. The second
number refers to the structure $Y_i$ of Table \ref{tab:nf1}.
$P_{57}=\lgl D_\mu\chi D^\mu\tilde\chi\rgl+$h.c.;
$\tilde\chi=\tau_2\chi^T\tau_2$.}
\endlastfoot
1(1)& $ 7/9$ & $  157/1728$ & $  19/12\,l^r_1 + 43/24\,l^r_2 + 1/2\,l^r_6 $\\

2(3)& $  - 11/48$ & $ - 1/24$ & $ - 1/4\,l^r_1 - 5/8\,l^r_2 - 1/8\,l^r_6 $\\

3(5)& $ 13/48$ & $  137/1728$ & $  1/3\,l^r_1 + 5/6\,l^r_2 + 3/8\,l^r_6 $\\

4(7)& $  - 31/36$ & $  503/864$ & $  7/6\,l^r_1 - 1/6\,l^r_2 - l^r_4 $\\

5(13)& $ 13/36$ & $ - 107/864$ & $  5/6\,l^r_1 - 5/6\,l^r_2 + 1/2\,l^r_4 $\\

6(17)& $ 43/72$ & $ - 7/864$ & $  11/12\,l^r_1 + 7/12\,l^r_2 + 1/4\,l^r_4 $\\

7(19)& & & $ 1/8\,l^r_7 $\\

8(20)& $ 5/16$ & $  1/8$ & $ 7/8\,l^r_1 + 1/2\,l^r_2 - 1/4\,l^r_3 
- 1/16\,l^r_4
 $\\

9(23)& & & $  - 1/8\,l^r_7 $\\

10(25)& $  - 3/64$ & & $ 3/16\,l^r_3 + 1/16\,l^r_7 $\\

11(26)& $ 9/128$ & & $ - 9/32\,l^r_3 - 1/32\,l^r_7 $\\

12(28)& $ 1/2$ & $ 7/432$ & $ 1/6\,l^r_1 + 2/3\,l^r_2 + 1/4\,l^r_4 $\\

13(31)& $  - 5/24$ & $ - 67/288$ & $ - l^r_1 - 2\,l^r_2 + 1/2\,l^r_4 - 3/4\,l^r_6
 $\\

14(33)& $ 239/288$ & $ 49/288$ & $ 5/3\,l^r_1 + 2\,l^r_2 - 1/8\,l^r_3 - 1/16
\,l^r_4 + 1/2\,l^r_6 $\\

15(34)& $  - 145/288$ & $ - 265/3456$ & $ - 7/6\,l^r_1 - 5/6\,l^r_2 - 1/16\,
l^r_4 - 3/16\,l^r_6 
 \extraline
 + 1/4\,l^r_7 $\\

16(37)& $ 17/96$ & $ - 29/1728$ & $ 2/3\,l^r_1 - 1/3\,l^r_2 + 1/8\,l^r_3 + 3/
16\,l^r_4 - 1/8\,l^r_6 $\\

17(39)& $ 145/576$ & $ 13/3456$ & $ 11/24\,l^r_1 + 7/24\,l^r_2 + 1/16\,l^r_3
 + 3/32\,l^r_4
\extraline
 - 5/16\,l^r_7 $\\

18(40)& $ 145/1152$ & $ 13/6912$ & $ 11/48\,l^r_1 + 7/48\,l^r_2 + 1/32\,
l^r_3 + 3/64\,l^r_4 
\extraline
+ 5/32\,l^r_7 $\\

19(41)& $  - 145/576$ & $ - 13/3456$ & $ - 11/24\,l^r_1 - 7/24\,l^r_2 - 1/16\,
l^r_3 - 3/32\,l^r_4
\extraline
 + 3/16\,l^r_7 $\\

20(43)& $  - 2/3$ & $ - 1/16$ & $ - l^r_1 - 3/4\,l^r_2 - 1/4\,l^r_4 $\\

21(44)& $ 2/3$ & $ 1/16$ & $ l^r_1 + 3/4\,l^r_2 + 1/4\,l^r_4 $\\

22(47) & & & \\

23(48)& $  - 1/16$ & $ 1/192$ & $ - 1/16\,l^r_4 $\\

24(49)& $  - 137/72$ & $ - 9/32$ & $ - 2\,l^r_1 - 16/3\,l^r_2 - 5/4\,l^r_6 $\\

25(54)& $ 5/36$ & $ - 67/432$ & $ 2\,l^r_1 - 1/3\,l^r_2 + 1/2\,l^r_6 $\\

26(58)& $ 55/72$ & $ 449/864$ & $ 8/3\,l^r_2 + 3/4\,l^r_6 $\\

27(66)& $  - 7/9$ & $ - 371/864$ & $ 1/6\,l^r_1 - 37/12\,l^r_2 - 4/3\,l^r_6 $\\

28(67)& $  - 97/36$ & $ - 245/288$ & $ - 25/6\,l^r_1 - 95/12\,l^r_2 - 23/6\,
l^r_6 $\\

29(71)& $  - 11/144$ & $ - 77/1728$ & $ - 1/6\,l^r_1 - 1/6\,l^r_2 - 3/2\,l^r_5
 + 17/24\,l^r_6 $\\

30(73)& $ 1/48$ & $ 85/1728$ & $ 1/6\,l^r_1 - 1/2\,l^r_5 + 7/24\,l^r_6 $\\

31(75)& $ 1/72$ & $ - 5/864$ & $ 1/2\,l^r_2 + 11/12\,l^r_6 $\\

32(76)& $ 5/18$ & $ - 11/216$ & $ 1/2\,l^r_1 + 1/4\,l^r_2 - 2/3\,l^r_6 $\\

33(78)& $  - 5/144$ & $ 3/64$ & $ - 1/4\,l^r_1 - 1/24\,l^r_2 - 1/8\,l^r_6 $\\

34(81)& & & $  - l^r_5 + 1/2\,l^r_6 $\\

35(85)& $  - 20/9$ & $ - 5/216$ & $ - 8/3\,l^r_1 - 17/6\,l^r_2 - l^r_4 - l^r_6 $\\

36(86)& $  - 5/24$ & $ 7/216$ & $ - 5/6\,l^r_1 + 1/12\,l^r_2 + 7/12\,l^r_6 $\\

38(89)& $  - 29/72$ & $ - 5/216$ & $ - 7/6\,l^r_1 - 5/12\,l^r_2 + 5/12\,l^r_6 $\\

39(90)& $  - 139/144$ & $ - 29/144$ & $ - 5/3\,l^r_1 - 7/4\,l^r_2 + 3/2\,l^r_5
 - 1/8\,l^r_6 $\\

40(92)& $ 1/4$ & $ 143/1728$ & $ 1/3\,l^r_1 + 7/12\,l^r_2 + 1/2\,l^r_5 - 1/4
\,l^r_6 $\\

41(94)& $ 59/72$ & $ 107/432$ & $ 5/6\,l^r_1 + 9/4\,l^r_2 + 5/12\,l^r_6 $\\

42(95)& $  - 7/6$ & $ - 137/864$ & $ - 8/3\,l^r_1 - 2\,l^r_2 + 1/3\,l^r_6 $\\

43(97)& $  - 17/24$ & $ - 11/96$ & $ - 17/12\,l^r_1 - 31/24\,l^r_2 + 1/4\,l^r_6
 $\\

44(100)& $  - 1/48$ & $ 41/1728$ & $ 1/12\,l^r_1 - 1/24\,l^r_2 + 1/8\,l^r_6 $\\

45(101)& $  - 1/16$ & $ - 23/1728$ & $ - 1/12\,l^r_1 + 1/24\,l^r_2 + 3/8\,l^r_6 $\\

46(102)& $  - 49/72$ & $ - 59/864$ & $ - 11/12\,l^r_1 - 7/12\,l^r_2 - 1/4\,l^r_4
 + l^r_5 $\\

47(104)& $  - 1/288$ & $ 85/3456$ & $ - 1/24\,l^r_1 + 1/48\,l^r_2 + 1/2\,l^r_5
 - 11/48\,l^r_6 $\\

48(105)& $ 127/144$ & $ 5/27$ & $ 3/2\,l^r_1 + 2\,l^r_2 + 5/24\,l^r_6 $\\

49(107)& $ 41/36$ & $ 19/432$ & $ 5/3\,l^r_1 + 5/6\,l^r_2 + 1/2\,l^r_4 - 1/2\,
l^r_6 $\\

50(109)& $  - 1/36$ & $ 1/96$ & $ 1/6\,l^r_6 $\\

51(110)& $  - 5/72$ & $ - 29/864$ & $ - 1/6\,l^r_1 + 1/12\,l^r_2 + 5/12\,l^r_6 $\\

52(111)& $  - 7/72$ & $ 1/72$ & $ 1/6\,l^r_1 - 1/12\,l^r_2 + 7/12\,l^r_6 $\\

53(112)& $  - 1/12$ & $ - 11/288$ & $ 1/2\,l^r_6 $\\

54(113)& $  - 1/2$ & $ - 1/48$ & $ - 1/2\,l^r_4 $\\

55(114)& $  - 2/9$ & & $ 4/3\,l^r_6 $\\

56(115)& $ 1/18$ & & $ - 1/3\,l^r_6 $\\

57& $  - 1/4$ & $ - 1/96$ & $ - 1/4\,l^r_4 $\\
\end{longtable}

\setcounter{equation}{0}
\addtocounter{zahler}{1}
\section{Divergences for $SU(n)$}
\label{app:su(n)}

\renewcommand{\extraline}{$\\ & & &$}

\begin{longtable}[c]{|r|l|c|c|}
\hline
$i$ & \hspace*{1.5cm}$Y_i$ & $\hat{\Gamma}^{(2)}_i$ & 
$16\pi^2\hat{\Gamma}^{(1)}_i$ \\
\hline
\hline
\endhead
\hline
\caption[]{\rule{0cm}{2em}}
\endfoot
\hline
\caption[Infinities]{\label{tab:nf1} 
\rule{0cm}{2em} 
The coefficients $\hat{\Gamma}^{(2)}_i$ of the double-pole
and $\hat{\Gamma}^{(1)}_i$ of the single-pole divergences
for the $n$-flavour case in Minkowski space.}
\endlastfoot
1 &$\lgl u\ccdot u h_{\mu \nu} h^{\mu\nu} \rgl$
 & $  7/16 + 5/288\,n^2$ & $ 5/96 - 5/6912\,n^2 $ \\

2 & $\lgl u\ccdot u \rgl \lgl h_{\mu \nu} h^{\mu\nu} \rgl$
 & $  41/576\,n $ & $ 11/3456\,n $ \\

3 &$\lgl h_{\mu \nu} u_\rho h^{\mu\nu} u^\rho \rgl$
 & $   - 3/16 $ & $  - 1/64 - 1/2304\,n^2 $ \\

4 & $\lgl h_{\mu \nu} u_\rho \rgl \lgl h^{\mu\nu} u^\rho \rgl$
 & $   - 1/72\,n $ & $ 1/864\,n $ \\

5 & $\lgl h_{\mu \nu} \big(u_\rho h^{\mu\rho} u^\nu $
  & $  1/4 $ & $ 7/192 + 5/6912\,n^2 $ \\
 & $ + u^\nu h^{\mu
    \rho} u_\rho \big) \rgl$ & & \\

6 & $\lgl h_{\mu \nu} u_\rho \rgl \lgl h^{\mu\rho} u^\nu \rgl$
 & $  1/96\,n $ & $ 23/1152\,n $ \\

7 &$\lgl (u\ccdot u)^2 \chi_+ \rgl$
 & $   - 1/36\,n^2 $ & $  - 13/3456\,n^2 $ \\

8 & $\lgl (u\ccdot u)^2 \rgl \lgl \chi_+ \rgl$
 & $   - 2/9\,n $ & $ 35/1728\,n $ \\

9 & $\lgl u\ccdot u \rgl \lgl u\ccdot u \chi_+ \rgl$
 & $  1/16\,n $ & $ 5/96\,n $ \\

10 & $\lgl u\ccdot u \rgl^2 \lgl \chi_+ \rgl$
 & $  1/32 $ & $ 5/128 $ \\

11 & $\lgl u\ccdot u u_\mu \chi_+ u^\mu \rgl$
 & $  3/8 - 17/288\,n^2 $ & $ 1/8 + 1/108\,n^2 $ \\

12& $\lgl u\ccdot u u_\mu \rgl \lgl \chi_+ u^\mu \rgl$
 &  & $  - 5/192\,n $\\

13 &$\lgl \chi_+ u_\mu u_\nu u^\mu u^\nu \rgl$
 & $   - 3/8 + 1/18\,n^2 $ & $  - 1/8 - 5/1728\,n^2 $ \\

14 & $\lgl \chi_+ \rgl \lgl  u_\mu u_\nu u^\mu u^\nu \rgl$
 & $  2/9\,n $ & $ 5/864\,n $ \\

15 & $\lgl \chi_+ u_\mu u_\nu \rgl \lgl u^\mu u^\nu \rgl$
 & $   - 1/8\,n $ & $  - 1/48\,n $ \\

16 & $\lgl \chi_+ \rgl \lgl u_\mu u_\nu \rgl^2$
 & $   - 1/16 $ & $ 1/64 $ \\

17 & $\lgl \chi_+ h_{\mu \nu} h^{\mu\nu} \rgl$
  & $  19/576\,n^2 $ & $  - 13/6912\,n^2 $ \\

18 & $\lgl \chi_+ \rgl \lgl h_{\mu \nu} h^{\mu\nu} \rgl$
 & $  67/576\,n $ & $  - 1/6912\,n $ \\

19 &$\lgl u\ccdot u \chi_+^2 \rgl$
 & $   - 3/32 $ & $  - 3/128 - 1/1536\,n^2$ \\ & & & $ + 1/32\,n^{-2} $ \\

20 &$\lgl u\ccdot u \chi_+ \rgl \lgl \chi_+ \rgl$
 & $  3/16\,n^{-1} $ & $ 5/192\,n + 1/64\,n^{-1} $ \\

21 & $\lgl u\ccdot u \rgl \lgl \chi_+^2 \rgl$
 & $  1/32\,n + 3/32\,n^{-1} $ & $ 5/384\,n - 1/128\,n^{-1} $ \\

22 & $\lgl u\ccdot u \rgl \lgl \chi_+ \rgl^2$
 & $  1/64 - 1/8\,n^{-2} $ & $ 5/256 $ \\

23 &$\lgl \chi_+ u_\mu \chi_+ u^\mu \rgl$
 & $  3/16 - 1/64\,n^2 $ & $  - 1/192 + 3/512\,n^2$ \\ & & & $ + 1/32\,n^{-2} $ \\

24 & $\lgl \chi_+ u_\mu \rgl^2$
 & $  1/64\,n - 5/16\,n^{-1} $ & $  - 3/64\,n^{-1} - 1/48\,n^{-3} $ \\

25 &$\lgl \chi_+^3 \rgl$
 & $   - 1/2\,n^{-2} $ & \\

26 &$\lgl \chi_+^2 \rgl \lgl \chi_+ \rgl$
 & $  1/4\,n^{-1} + 1/2\,n^{-3} $ & \\

27 & $\lgl \chi_+ \rgl^3$
 & $   - 1/8\,n^{-2} - 1/8\,n^{-4} $ & \\

28 &i $\lgl \chi_- \{ h_{\mu \nu},u^\mu u^\nu\} \rgl$
 & $  1/48 - 1/192\,n^2 $ & $ 17/1152 + 1/432\,n^2 $ \\

29 & i $\lgl \chi_-  h_{\mu \nu} \rgl \lgl u^\mu u^\nu \rgl$
 & $   - 1/96\,n $ & $ 1/128\,n $ \\

30 & i $\lgl h_{\mu \nu} u^\mu u^\nu \rgl \lgl \chi_- \rgl$
 & $  7/96\,n + 1/2\,n^{-1} $ & $ 41/3456\,n + 7/96\,n^{-1} $ \\

31 &i $\lgl h_{\mu \nu} u^\mu \chi_- u^\nu \rgl$
 & $   - 13/24 $ & $  - 59/576 - 1/1152\,n^2 $ \\

32 & i $\lgl h_{\mu \nu} u^\mu \rgl \lgl \chi_- u^\nu \rgl$
 & $   - 5/32\,n $ & $  - 25/576\,n $ \\

33 &$\lgl u\ccdot u \chi_-^2 \rgl$
 & $  5/12 - 1/8\,n^{-2} $ & $ 25/288 + 1/768\,n^2 $ \\
 & &  $+ 19/576\,n^2 $ & \\

34 &$\lgl u\ccdot u \chi_- \rgl \lgl \chi_- \rgl$
 & $   - 49/288\,n - 1/4\,n^{-1} $ & $  - 7/864\,n - 1/24\,n^{-1}$ \\
 & & & $ - 1/24\,n^{-3} $ \\

35 & $\lgl u\ccdot u \rgl \lgl \chi_-^2 \rgl$
 & $  1/18\,n + 1/16\,n^{-1} $ & $ 13/6912\,n + 1/32\,n^{-1} $ \\

36 & $\lgl u\ccdot u \rgl \lgl \chi_- \rgl^2$
 & $   - 31/576 + 1/4\,n^{-2} $ & $  - 23/1152 + 7/192\,n^{-2} $ \\

37 &$\lgl u_\mu \chi_- u^\mu \chi_- \rgl$
 & $   - 5/48 + 1/24\,n^2 $ & $ - 17/576-1/864\,n^2 $\\
 & & $ + 1/8\,n^{-2} $ & $+ 1/8\,n^{-2} $ \\

38 & $\lgl u_\mu \chi_- \rgl^2$
 & $  49/144\,n - 1/4\,n^{-1} $ & $ 103/3456\,n - 3/32\,n^{-1} $ \\

39 &$\lgl \chi_-^2 \chi_+ \rgl$
 & $  1/8 - 5/8\,n^{-2} $ & $ 1/96 
  + 5/6912\,n^2$ \\
 & & $+ 5/288\,n^2 $ &$ - 1/4\,n^{-2} $ \\

40 &$\lgl \chi_+ \rgl \lgl \chi_-^2 \rgl$
 & $  29/288\,n + 1/4\,n^{-3} $ & $  - 5/3456\,n 
  + 1/32\,n^{-1}$\\ & & & $ + 1/8\,n^{-3} $ \\

41 &$\lgl \chi_+ \chi_- \rgl \lgl \chi_- \rgl $
 & $   - 7/72\,n + 1/4\,n^{-1} $ & $  
     - 23/3456\,n + 11/96\,n^{-1} $ \\
 & & $+ 1/4\,n^{-3}$ &  $+ 1/24\,n^{-3}$\\

42 & $\lgl \chi_+ \rgl \lgl \chi_- \rgl^2$
 & $   - 1/12 - 1/16\,n^{-2} $ & $  - 55/2304 - 1/24\,n^{-4} $ \\
 & & $  - 1/8\,n^{-4} $ & \\

43 &i $\lgl \chi_- \{ \chi_{+ \, \mu},u^\mu \} \rgl$
 & $  1/16 - 5/96\,n^2 $ & $ 1/192 - 1/768\,n^2$\\ & & & $ - 1/16\,n^{-2} $ \\

44 &i $\lgl \chi_- \rgl \lgl \chi_{+ \, \mu} u^\mu \rgl$
 & $  5/48\,n - 1/8\,n^{-1} $ & $  - 1/384\,n + 1/32\,n^{-1} 
$\\ & & & $ + 1/24\,n^{-3} $ \\

45 & i $\lgl \chi_{+ \mu} \rgl \lgl \chi_- u^\mu \rgl$
 & $   - 5/12\,n + 1/8\,n^{-1} $ & $  - 1/32\,n + 1/32\,n^{-1} $ \\

46 & $\lgl \chi_{- \mu} \rgl^2$
 & & $ 5/768\,n - 1/32\,n^{-1}$\\ & & & $ - 1/48\,n^{-3} $ \\

47 &$\lgl \chi_{+ \mu}  \chi_+^\mu \rgl$
 & $  1/8 - 1/32\,n^2 $ & $  - 1/48 + 1/256\,n^2$\\ & & & $ + 1/16\,n^{-2} $ \\

48 &$\lgl \chi_{+ \mu} \rgl^2$
 & $   - 3/32\,n $ &\\

49 &$\lgl (u\ccdot u)^3 \rgl$
 & $   - 1/3 - 1/96\,n^2 $ & $  - 37/576 - 5/1728\,n^2 $\\

50 & $\lgl (u\ccdot u)^2 \rgl \lgl u\ccdot u \rgl$
 & $   - 31/288\,n $ & $ 41/1728\,n $ \\

51 & $\lgl u\ccdot u \rgl^3$
 & $  1/64 $ & $ 5/256 $ \\

52 & $\lgl u\ccdot u u_\mu u\ccdot u u^\mu \rgl$
 & $   - 1/6 - 17/576\,n^2 $ & $ 7/144 + 1/256\,n^2 $ \\

53 & $\lgl u\ccdot u u_\mu \rgl^2$
 & $   - 1/64\,n $ & $  - 5/256\,n $ \\

54 &$\lgl u\ccdot u u_\mu u_\nu u^\mu u^\nu \rgl$
 & $  2/3 + 7/288\,n^2 $ & $ 41/576 + 5/1728\,n^2 $ \\

55 & $\lgl u\ccdot u u_\mu u_\nu \rgl \lgl u^\mu u^\nu \rgl$
 & $   - 1/24\,n $ & $  - 1/72\,n $ \\

56 & $\lgl u\ccdot u \rgl \lgl u_\mu u_\nu \rgl^2$
 & $   - 1/32 $ & $ 3/128 $ \\

57 & $\lgl u\ccdot u \rgl \lgl u_\mu u_\nu u^\mu u^\nu \rgl$
 & $  5/36\,n $ & $ 53/3456\,n $ \\

58 &$\lgl u_\mu u_\nu u_\rho u^\mu u^\nu u^\rho \rgl$
 & $  1/6 + 1/576\,n^2 $ & $ 65/576 + 5/6912\,n^2 $ \\

59 & $\lgl u_\mu u_\nu u_\rho \rgl^2$
 & $   - 1/192\,n $ & $ 65/2304\,n $ \\

60 & $\lgl u_\mu u_\nu u_\rho u^\mu u^\rho u^\nu \rgl$
 & $   - 1/3 - 1/576\,n^2 $ & $  - 97/576 - 23/6912\,n^2 $ \\

61 & $\lgl u_\mu u_\nu u_\rho \rgl \lgl u^\mu u^\rho u^\nu \rgl$
 & $  7/192\,n $ & $  - 23/2304\,n $ \\

62 & $\lgl u_\mu u_\nu \rgl \lgl u_\rho u^\mu u^\rho u^\nu \rgl$
 & $   - 1/12\,n $ & $  - 5/288\,n $ \\

63 & $\lgl u_\mu u_\nu \rgl \lgl u^\mu u_\rho \rgl \lgl u^\nu u^\rho \rgl$
 & $   - 1/8 $ & $  - 1/32 $ \\

64 & i $\lgl f_{+ \mu \nu} \{u\ccdot u, u^\mu u^\nu \} \rgl$
 & $   - 1/4 - 1/64\,n^2 $ & $  - 11/192 - 1/864\,n^2 $ \\

65 & i $\lgl u\ccdot u \rgl \lgl f_{+ \mu \nu} u^\mu u^\nu \rgl$
 & $   - 7/24\,n $ & $  - 7/192\,n $ \\

66 & i $\lgl f_{+ \mu \nu} u_\rho  u^\mu u^\nu u^\rho \rgl$
 & $   - 3/8 - 1/48\,n^2 $ & $  - 7/96 + 5/1728\,n^2 $ \\

67 &i $\lgl f_{+ \mu \nu} u^\mu u\ccdot u u^\nu \rgl$ & $   - 7/8 - 1/32\,n^2 $ & $  - 3/16 + 1/432\,n^2 $ \\

68 & i $\lgl f_{+ \mu \nu} \{ u_\rho , u^\mu u^\rho u^\nu \} \rgl$ & $  3/8 + 1/72\,n^2 $ & $ 11/96 + 1/3456\,n^2 $ \\

69 & i $\lgl f_{+ \mu \nu} u_\rho \rgl \lgl  u^\mu u^\nu u^\rho \rgl$
 & $  1/48\,n $ & $  - 47/576\,n $ \\

70 & i $\lgl f_{+ \mu \nu} [u^\mu ,u_\rho ] \rgl \lgl u^\nu u^\rho \rgl$
 & $  1/9\,n $ & $ 1/216\,n $ \\

71 &$\lgl u\ccdot u f_{+ \mu \nu} f_+^{\mu\nu} \rgl$
 & $  1/72\,n^2 $ & $  - 1/48 - 1/288\,n^2 $ \\

72 & $\lgl u\ccdot u \rgl \lgl f_{+ \mu \nu} f_+^{\mu\nu} \rgl$
 & $   - 1/72\,n $ & $  - 5/864\,n $ \\

73 &$\lgl f_{+ \mu \nu} u_\rho f_+^{\mu\nu} u^\rho \rgl$
 & $   - 5/288\,n^2 $ & $ 1/48 + 13/3456\,n^2 $ \\

74 & $\lgl f_{+ \mu \nu} u_\rho \rgl^2$
 & $  1/288\,n $ & $ 23/3456\,n $ \\

75 &$\lgl f_{+ \mu \nu} f_+^{\mu\rho} u^\nu u_\rho \rgl$
 & $  1/12 $ & $  - 1/72 - 1/864\,n^2 $ \\

76 & $\lgl f_{+ \mu \nu} f_+^{\mu\rho} u_\rho u^\nu \rgl$
 & $   - 1/24 - 1/36\,n^2 $ & $  - 7/288 + 5/864\,n^2 $ \\

77 & $\lgl f_{+ \mu \nu} f_+^{\mu\rho} \rgl \lgl u^\nu u_\rho \rgl$ 
 & $  1/18\,n $ & $  - 11/432\,n $ \\

78 &$\lgl f_{+ \mu \nu} \big( u_\rho f_+^{\mu\rho} u^\nu$
 & $   - 1/48 + 1/48\,n^2 $ & $ 11/576 - 1/864\,n^2 $ \\
 & $+ u^\nu f_+^{
    \mu \rho}  u_\rho \big) \rgl$ & & \\

79 & $\lgl f_{+ \mu \nu} u_\rho \rgl \lgl f_+^{\mu\rho} u^\nu \rgl$
 & $   - 1/144\,n $ & $ 55/1728\,n $ \\

80 & $\lgl f_{+ \mu \nu} u^\nu \rgl \lgl f_+^{\mu\rho} u_\rho \rgl$
 & $   - 1/144\,n $ & $ 1/1728\,n $ \\

81 &$\lgl \chi_+ f_{+ \mu \nu} f_+^{\mu\nu} \rgl$ & & \\

82 & $\lgl \chi_+ \rgl \lgl f_{+ \mu \nu} f_+^{\mu\nu} \rgl$ & & \\

83 & i $\lgl f_{+ \mu \nu} \{\chi_+, u^\mu u^\nu \} \rgl$
 & $   - 1/36\,n^2 $ & $ 5/3456\,n^2 $ \\

84 & i $\lgl \chi_+ \rgl \lgl f_{+ \mu \nu} u^\mu u^\nu \rgl$
 & $   - 4/9\,n $ & $  - 5/432\,n $ \\

85 &i $\lgl f_{+ \mu \nu} u^\mu \chi_+ u^\nu \rgl$
 & $   - 1/18\,n^2 $ & $ 5/1728\,n^2 $ \\

86 &$\lgl f_{- \mu \nu} \big(h^{\nu\rho} u_\rho u^\mu $
 & $  1/12 - 5/288\,n^2 $ & $ 7/576 + 1/1152\,n^2 $ \\
  & $+ u^\mu u_\rho
   h^{\nu\rho} \big) \rgl$  & & \\

87 &$\lgl f_{- \mu \nu} h^{\nu\rho} \rgl \lgl u^\mu u_\rho \rgl$
 & $  1/144\,n $ & $ 23/1728\,n $ \\

88 & $\lgl f_{- \mu \nu} u^\mu \rgl \lgl h^{\nu\rho} u_\rho \rgl$
 & $   - 17/72\,n $ & $  - 17/1728\,n $ \\

89 &$\lgl f_{- \mu \nu} \big(u^\mu h^{\nu\rho} u_\rho$
 & $   - 1/12 - 1/48\,n^2 $ & $  - 7/576 - 1/3456\,n^2 $ \\
 & $+  u_\rho
   h^{\nu\rho} u^\mu \big) \rgl$ & & \\

90 & $\lgl u\ccdot u f_{- \mu \nu} f_-^{\mu\nu} \rgl$
 & $   - 5/12 - 1/36\,n^2 $ & $  - 47/576 - 13/2304\,n^2 $ \\

91 & $\lgl u\ccdot u \rgl \lgl f_{- \mu \nu} f_-^{\mu\nu} \rgl$
 & $   - 47/576\,n $ & $  - 31/1728\,n $ \\

92 & $\lgl f_{- \mu \nu} u_\rho f_-^{\mu\nu} u^\rho \rgl$
 & $  1/6 $ & $ 13/288 + 17/6912\,n^2 $ \\

93 &$\lgl f_{- \mu \nu} u_\rho \rgl^2$
 & $   - 1/144\,n $ & $ 1/1728\,n $ \\

94 &$\lgl f_{- \mu \nu} f_-^{\mu\rho} u^\nu u_\rho \rgl$
 & $  5/8 - 1/144\,n^2 $ & $ 11/96 + 23/3456\,n^2 $ \\

95 &$\lgl f_{- \mu \nu} f_-^{\mu\rho} u_\rho u^\nu \rgl$
 & $   - 1/2 - 1/24\,n^2 $ & $  - 7/96 - 11/3456\,n^2 $ \\

96 & $\lgl f_{- \mu \nu} f_-^{\mu\rho} \rgl \lgl u^\nu u_\rho \rgl$
 & $  1/144\,n $ & $ 23/1728\,n $ \\

97 &$\lgl f_{- \mu \nu} \big( u_\rho f_-^{\mu\rho} u^\nu$
 & $   - 5/16 - 1/72\,n^2 $ & $  - 11/192 - 13/6912\,n^2 $ \\
 & $ + u^\nu f_-^{
    \mu \rho}  u_\rho \big) \rgl$ & & \\

98 &$\lgl f_{- \mu \nu} u_\rho \rgl \lgl f_-^{\mu\rho} u^\nu \rgl$
 & $  1/288\,n $ & $ 23/3456\,n $ \\

99 &$\lgl f_{- \mu \nu} u^\nu \rgl \lgl f_-^{\mu\rho} u_\rho \rgl$
 & $   - 1/4\,n $ & $  - 7/192\,n $ \\

100 &i $\lgl f_{+ \mu \nu} [f_-^{ \nu \rho},h^\mu_\rho] \rgl$
 & $   - 1/48 - 1/48\,n^2 $ & $ 5/576 + 13/3456\,n^2 $ \\

101 &i $\lgl f_{+ \mu \nu} [f_-^{ \nu \rho},f_{-\rho}^\mu] \rgl$
 & $  1/48 $ & $  - 5/576 - 1/864\,n^2 $ \\

102 & $\lgl \chi_+ f_{- \mu \nu} f_-^{\mu\nu} \rgl$
 & $   - 25/576\,n^2 $ & $  - 17/6912\,n^2 $ \\

103 & $\lgl \chi_+ \rgl \lgl f_{- \mu \nu} f_-^{\mu\nu} \rgl$
 & $   - 73/576\,n $ & $  - 101/6912\,n $ \\

104 &$\lgl f_{+ \mu \nu} [f_-^{\mu\nu},\chi_- ] \rgl$
 & $  1/96 + 1/144\,n^2 $ & $ 19/1152 + 7/3456\,n^2 $ \\

105 &i $\lgl f_{- \mu \nu} [\chi_-, u^\mu u^\nu ] \rgl$
 & $  1/4 + 7/192\,n^2 $ & $ 31/384 + 1/256\,n^2 $ \\

106 & i $\lgl f_{- \mu \nu} u^\nu \rgl \lgl u^\mu \chi_- \rgl$
 & $  53/96\,n $ & $ 1/18\,n $ \\

107 & $\lgl f_{- \mu \nu} \{ \chi_+^\mu, u^\nu \} \rgl$
 & $  5/72\,n^2 $ & $  - 1/1728\,n^2 $ \\

108 & $\lgl \chi_+^\mu \rgl \lgl f_{- \mu \nu} u^\nu \rgl$
 & $  31/72\,n $ & $ 5/216\,n $ \\

109 &$\lgl \nabla_\rho f_{- \mu \nu} \nabla^\rho f_-^{\mu\nu} \rgl$
 & $   - 1/144\,n^2 $ & $ 1/384\,n^2 $ \\

110 &i $\lgl \nabla_\rho f_{+ \mu \nu} [h^{\mu\rho}, u^\nu ] \rgl$
 & $  1/24 + 1/72\,n^2 $ & $  - 5/288 - 7/1728\,n^2 $ \\

111 &i $\lgl \nabla^\mu f_{+ \mu \nu} [f_-^{ \nu \rho}, u_\rho ]
\rgl$ & $   - 7/288\,n^2 $ & $ 1/288\,n^2 $ \\

112 &i $\lgl \nabla^\mu f_{+ \mu \nu} [h^{\nu\rho}, u_\rho ] \rgl$
 & $  1/24 + 1/96\,n^2 $ & $  - 5/288 - 1/192\,n^2 $ \\

113 &$\lgl D_\mu \chi D^\mu \chi^\dagger \rgl$
 & & $ 1/12 - 1/64\,n^2 - 1/4\,n^{-2} $\\

114 &i $\lgl F_{L \mu \nu} F_L^{ \mu \rho} F_{L \rho}^\nu \rgl $
 & $   - 1/18\,n^2 $ &  \\
 &  +$L \rightarrow R$ & & \\

115 &$\lgl D_\rho F_{L \mu \nu} D^\rho F_L^{ \mu \nu} \rgl$
  & $  1/72\,n^2 $ & \\
 &  +$L \rightarrow R$ & & \\
\end{longtable}


\renewcommand{\extraline}{$\\&$}
\begin{longtable}[c]{|r|l|}
\hline
$Y_i$  & \hspace*{5cm} $\hat{\Gamma}^{(L)}_i$\\
\hline
\hline
\endhead
\hline
\caption[]{\rule{0cm}{2em}}
\endfoot
\hline
\caption[Infinities]{\label{tab:nf2} 
\rule{0cm}{2em} The coefficients $\hat{\Gamma}^{(L)}_i$ of the single-pole
divergences for the $n$-flavour case in Minkowski space.}
\endlastfoot
1& $  1/6\,n\,\hat{L}^r_{0} + 3/4\,n\,\hat{L}^r_{3} + 7/3\,\hat{L}^r_{1} + 35/6\,\hat{L}^r_{2} $ \\

2& $  n\,\hat{L}^r_{1} + 1/3\,n\,\hat{L}^r_{2} + 5/6\,\hat{L}^r_{0} + 3/4\,\hat{L}^r_{3} - 1/8\,\hat{L}^r_{9} $ \\

3& $   - \hat{L}^r_{1} - 5/2\,\hat{L}^r_{2} $ \\

4& $   - 1/6\,\hat{L}^r_{3} - 1/4\,\hat{L}^r_{9} $ \\

5& $   - 1/6\,n\,\hat{L}^r_{0} + 1/12\,n\,\hat{L}^r_{3} + 4/3\,\hat{L}^r_{1} + 10/3\,\hat{L}^r_{2} $ \\

6& $  3\,\hat{L}^r_{0} + 1/2\,\hat{L}^r_{3} - 3/4\,\hat{L}^r_{9} $ \\

7& $  1/6\,n\,\hat{L}^r_{0} + 1/12\,n\,\hat{L}^r_{3} - 1/2\,n\,\hat{L}^r_{5} - 8/3\,n^{-1}\,\hat{L}^r_{0}
 - 8/3\,n^{-1}\,\hat{L}^r_{3} + 4/3\,\hat{L}^r_{1}
\extraline
 + 2/3\,\hat{L}^r_{2} $ \\

8& $   - 2\,n\,\hat{L}^r_{1} - 2/3\,n\,\hat{L}^r_{2} - n\,\hat{L}^r_{4} + 2\,n^{-1}\,\hat{L}^r_{1} + 2/3\,
n^{-1}\,\hat{L}^r_{2} - 10/3\,n^{-2}\,\hat{L}^r_{0}
\extraline
 - 10/3\,n^{-2}\,\hat{L}^r_{3} - 2/3\,\hat{L}^r_{0} - 5/6\,\hat{L}^r_{3} - 1/2\,
\hat{L}^r_{5} $ \\

9& $  n\,\hat{L}^r_{1} + 1/2\,n\,\hat{L}^r_{2} - 1/2\,n\,\hat{L}^r_{4} + \hat{L}^r_{0} + 5/2\,\hat{L}^r_{3} - 1/
2\,\hat{L}^r_{5} $ \\

10& $  \hat{L}^r_{1} + 1/2\,\hat{L}^r_{2} - 1/2\,\hat{L}^r_{4} $ \\

11& $   - 5/6\,n\,\hat{L}^r_{0} - 11/12\,n\,\hat{L}^r_{3} - 1/2\,n\,\hat{L}^r_{5} + 28/3\,
n^{-1}\,\hat{L}^r_{0} + 28/3\,n^{-1}\,\hat{L}^r_{3}
\extraline
 + 4/3\,\hat{L}^r_{1} + 2/3\,\hat{L}^r_{2} $ \\

12& $  2\,\hat{L}^r_{0} - \hat{L}^r_{3} $ \\

13& $  1/6\,n\,\hat{L}^r_{0} + 13/12\,n\,\hat{L}^r_{3} + 1/2\,n\,\hat{L}^r_{5} - 20/3\,n^{-1}\,
\hat{L}^r_{0} - 20/3\,n^{-1}\,\hat{L}^r_{3}\extraline
 - 8/3\,\hat{L}^r_{1} - 4/3\,\hat{L}^r_{2} $ \\

14& $  2\,n\,\hat{L}^r_{1} + 2/3\,n\,\hat{L}^r_{2} + n\,\hat{L}^r_{4} - 2\,n^{-1}\,\hat{L}^r_{1} - 2/3\,
n^{-1}\,\hat{L}^r_{2} + 10/3\,n^{-2}\,\hat{L}^r_{0}\extraline
 + 10/3\,n^{-2}\,\hat{L}^r_{3} + 1/6\,\hat{L}^r_{0} + 13/12\,\hat{L}^r_{3} + 1/2\,
\hat{L}^r_{5} $ \\

15& $   - n\,\hat{L}^r_{2} - 4\,\hat{L}^r_{0} - \hat{L}^r_{3} $ \\

16& $   - \hat{L}^r_{2} $ \\

17& $  1/3\,n\,\hat{L}^r_{0} + 2/3\,n\,\hat{L}^r_{3} + 1/4\,n\,\hat{L}^r_{5} - 10/3\,n^{-1}\,\hat{L}^r_{0}
 - 10/3\,n^{-1}\,\hat{L}^r_{3} + 2/3\,\hat{L}^r_{1}\extraline
 + 4/3\,\hat{L}^r_{2} $ \\

18& $  n\,\hat{L}^r_{1} + 1/3\,n\,\hat{L}^r_{2} + 1/2\,n\,\hat{L}^r_{4} - n^{-1}\,\hat{L}^r_{1} - 1/3\,
n^{-1}\,\hat{L}^r_{2} + 5/3\,n^{-2}\,\hat{L}^r_{0}\extraline
 + 5/3\,n^{-2}\,\hat{L}^r_{3} + 1/3\,\hat{L}^r_{0} + 2/3\,\hat{L}^r_{3} + 1/4\,\hat{L}^r_{5}
 $ \\

19& $  1/4\,n\,\hat{L}^r_{0} + 5/8\,n\,\hat{L}^r_{3} - 1/2\,n\,\hat{L}^r_{8} - 7\,n^{-1}\,\hat{L}^r_{0} - 
17/2\,n^{-1}\,\hat{L}^r_{3} + 1/2\,\hat{L}^r_{1} \extraline
+ 5/4\,\hat{L}^r_{2} - 2\,\hat{L}^r_{7} $ \\

20& $  1/2\,n\,\hat{L}^r_{4} - n\,\hat{L}^r_{6} + 6\,n^{-2}\,\hat{L}^r_{0} + 6\,n^{-2}\,\hat{L}^r_{3} - 
n^{-2}\,\hat{L}^r_{5} + 1/2\,\hat{L}^r_{0} + 5/4\,\hat{L}^r_{3} - \hat{L}^r_{8} $ \\

21& $  n\,\hat{L}^r_{1} + 1/4\,n\,\hat{L}^r_{2} - 1/2\,n\,\hat{L}^r_{4} - 4\,n^{-1}\,\hat{L}^r_{1} - n^{-1}
\,\hat{L}^r_{2} + 2\,n^{-1}\,\hat{L}^r_{4}\extraline
 + 3\,n^{-2}\,\hat{L}^r_{0} + 3\,n^{-2}\,\hat{L}^r_{3} + 1/4\,\hat{L}^r_{0} + 5/8\,\hat{L}^r_{3} + 1/4
\,\hat{L}^r_{5} - 1/2\,\hat{L}^r_{8} $ \\

22& $  2\,n^{-2}\,\hat{L}^r_{1} + 1/2\,n^{-2}\,\hat{L}^r_{2} - n^{-2}\,\hat{L}^r_{4} - 3\,n^{-3}\,\hat{L}^r_{0}
 - 3\,n^{-3}\,\hat{L}^r_{3} + \hat{L}^r_{1} + 1/4\,\hat{L}^r_{2} 
\extraline
- \hat{L}^r_{6} $ \\

23& $   - 1/4\,n\,\hat{L}^r_{5} - 2\,n^{-1}\,\hat{L}^r_{0} - 1/2\,n^{-1}\,\hat{L}^r_{3} + 2\,n^{-1}
\,\hat{L}^r_{5} + 1/2\,\hat{L}^r_{1} + 5/4\,\hat{L}^r_{2} + 2\,\hat{L}^r_{7} $ \\

24& $   - n^{-1}\,\hat{L}^r_{1} - 5/2\,n^{-1}\,\hat{L}^r_{2} + \hat{L}^r_{0} + 1/4\,\hat{L}^r_{3} - 1/2\,
\hat{L}^r_{5} + \hat{L}^r_{8} $ \\

25& $  1/4\,n\,\hat{L}^r_{5} - 1/2\,n\,\hat{L}^r_{8} - 3\,n^{-1}\,\hat{L}^r_{5} + 4\,n^{-1}\,\hat{L}^r_{8}
 - 2\,\hat{L}^r_{7} $ \\

26& $  1/2\,n\,\hat{L}^r_{4} - n\,\hat{L}^r_{6} - 2\,n^{-1}\,\hat{L}^r_{4} + 4\,n^{-1}\,\hat{L}^r_{6} + 4\,
n^{-1}\,\hat{L}^r_{7} + 3\,n^{-2}\,\hat{L}^r_{5} - 2\,n^{-2}\,\hat{L}^r_{8}\extraline
 + 3/4\,\hat{L}^r_{5} - 3/2\,\hat{L}^r_{8} $ \\

27& $  n^{-2}\,\hat{L}^r_{4} - 2\,n^{-2}\,\hat{L}^r_{6} - 2\,n^{-2}\,\hat{L}^r_{7} - n^{-3}\,\hat{L}^r_{5} + 1/
2\,\hat{L}^r_{4} - \hat{L}^r_{6} $ \\

28& $   - 1/3\,n\,\hat{L}^r_{0} - 1/3\,n\,\hat{L}^r_{3} + 1/12\,n\,\hat{L}^r_{5} + 2/3\,\hat{L}^r_{1}
 + 2/3\,\hat{L}^r_{2} - 2/3\,\hat{L}^r_{4} $ \\

29& $  \hat{L}^r_{0} + 1/2\,\hat{L}^r_{3} - 1/2\,\hat{L}^r_{5} $ \\

30& $  8/3\,n^{-1}\,\hat{L}^r_{1} + 20/3\,n^{-1}\,\hat{L}^r_{2} + 2/3\,\hat{L}^r_{0} + 5/3\,\hat{L}^r_{3}
 + 1/2\,\hat{L}^r_{5} $ \\

31& $   - n\,\hat{L}^r_{3} + 1/3\,n\,\hat{L}^r_{5} - 4\,\hat{L}^r_{1} - 8\,\hat{L}^r_{2} + 4/3\,\hat{L}^r_{4} $ \\

32& $   - 2\,n\,\hat{L}^r_{1} - n\,\hat{L}^r_{2} - 3\,\hat{L}^r_{0} - 3/2\,\hat{L}^r_{3} + 3/4\,\hat{L}^r_{9} $ \\

33& $  1/2\,n\,\hat{L}^r_{0} + 19/12\,n\,\hat{L}^r_{3} + 1/6\,n\,\hat{L}^r_{5} - 1/2\,n\,\hat{L}^r_{8}
 - 4\,n^{-1}\,\hat{L}^r_{0} - 4\,n^{-1}\,\hat{L}^r_{3}\extraline
 + n^{-1}\,\hat{L}^r_{5} + 11/3\,\hat{L}^r_{1} + 43/6\,\hat{L}^r_{2} - 4/3\,\hat{L}^r_{4}
 - 2\,\hat{L}^r_{6} $ \\

34& $   - n\,\hat{L}^r_{7} - 14/3\,n^{-1}\,\hat{L}^r_{1} - 29/3\,n^{-1}\,\hat{L}^r_{2} + 2\,n^{-1}
\,\hat{L}^r_{4} + 8\,n^{-2}\,\hat{L}^r_{0} + 8\,n^{-2}\,\hat{L}^r_{3}\extraline
 - 4/3\,\hat{L}^r_{0} - 3\,\hat{L}^r_{3} - \hat{L}^r_{5} - \hat{L}^r_{8} $ \\

35& $  n\,\hat{L}^r_{1} + 1/3\,n\,\hat{L}^r_{2} + 5/6\,\hat{L}^r_{0} + 3/4\,\hat{L}^r_{3} - 1/2\,\hat{L}^r_{8}
 - 1/8\,\hat{L}^r_{9} $ \\

36& $   - 2/3\,n^{-1}\,\hat{L}^r_{0} + 1/8\,n^{-1}\,\hat{L}^r_{9} + 4/3\,n^{-2}\,\hat{L}^r_{1} + 
10/3\,n^{-2}\,\hat{L}^r_{2} - \hat{L}^r_{1} - 1/3\,\hat{L}^r_{2}\extraline
 - \hat{L}^r_{7} $ \\

37& $  2/3\,n\,\hat{L}^r_{0} + 2/3\,n\,\hat{L}^r_{3} + 1/3\,n\,\hat{L}^r_{5} - 4\,n^{-1}\,\hat{L}^r_{0} - 
4\,n^{-1}\,\hat{L}^r_{3} + n^{-1}\,\hat{L}^r_{5} - 1/3\,\hat{L}^r_{1}\extraline
 - 5/6\,\hat{L}^r_{2} - 2/3\,\hat{L}^r_{4} + 2\,\hat{L}^r_{6} $ \\

38& $  2\,n\,\hat{L}^r_{1} + n\,\hat{L}^r_{2} + 2\,n\,\hat{L}^r_{4} - 2\,n^{-1}\,\hat{L}^r_{1} - n^{-1}\,\hat{L}^r_{2}
 - n^{-1}\,\hat{L}^r_{4} + 4\,n^{-2}\,\hat{L}^r_{0} \extraline
+ 4\,n^{-2}\,\hat{L}^r_{3} - n^{-2}\,\hat{L}^r_{5} + \hat{L}^r_{0} + 4/3\,\hat{L}^r_{3} + 1/
2\,\hat{L}^r_{5} + \hat{L}^r_{8} - 1/4\,\hat{L}^r_{9} $ \\

39& $  1/3\,n\,\hat{L}^r_{0} + 2/3\,n\,\hat{L}^r_{3} + 1/4\,n\,\hat{L}^r_{5} - 1/2\,n\,\hat{L}^r_{8} - 
10/3\,n^{-1}\,\hat{L}^r_{0} - 10/3\,n^{-1}\,\hat{L}^r_{3}\extraline
 + 4\,n^{-1}\,\hat{L}^r_{8} + 2/3\,\hat{L}^r_{1} + 4/3\,\hat{L}^r_{2} - 2\,
\hat{L}^r_{6} $ \\

40& $  n\,\hat{L}^r_{1} + 1/3\,n\,\hat{L}^r_{2} + 1/2\,n\,\hat{L}^r_{4} - n^{-1}\,\hat{L}^r_{1} - 1/3\,
n^{-1}\,\hat{L}^r_{2} + 5/3\,n^{-2}\,\hat{L}^r_{0}\extraline
 + 5/3\,n^{-2}\,\hat{L}^r_{3} - 2\,n^{-2}\,\hat{L}^r_{8} + 1/3\,\hat{L}^r_{0} + 2/3\,
\hat{L}^r_{3} + 1/4\,\hat{L}^r_{5} - 1/2\,\hat{L}^r_{8} $ \\

41& $   - n\,\hat{L}^r_{7} - 4/3\,n^{-1}\,\hat{L}^r_{1} - 8/3\,n^{-1}\,\hat{L}^r_{2} + 4\,n^{-1}\,
\hat{L}^r_{6} + 4\,n^{-1}\,\hat{L}^r_{7} + 20/3\,n^{-2}\,\hat{L}^r_{0}\extraline
 + 20/3\,n^{-2}\,\hat{L}^r_{3} - 2/3\,\hat{L}^r_{0} - 4/3\,\hat{L}^r_{3}
 - 1/2\,\hat{L}^r_{5} - \hat{L}^r_{8} $ \\

42& $  5/3\,n^{-2}\,\hat{L}^r_{1} + 5/3\,n^{-2}\,\hat{L}^r_{2} - 2\,n^{-2}\,\hat{L}^r_{6} - 2\,n^{-2}
\,\hat{L}^r_{7} - 5\,n^{-3}\,\hat{L}^r_{0} - 5\,n^{-3}\,\hat{L}^r_{3}\extraline
 - \hat{L}^r_{1} - 1/3\,\hat{L}^r_{2} - 1/2\,\hat{L}^r_{4} - \hat{L}^r_{7} $ \\

43& $   - 1/2\,n\,\hat{L}^r_{0} - 3/4\,n\,\hat{L}^r_{3} - 1/2\,n\,\hat{L}^r_{5} + 4\,n^{-1}\,\hat{L}^r_{0}
 + 4\,n^{-1}\,\hat{L}^r_{3} - \hat{L}^r_{1} - 3/2\,\hat{L}^r_{2} 
\extraline+ \hat{L}^r_{4} $ \\

44& $  2\,n^{-1}\,\hat{L}^r_{1} + 3\,n^{-1}\,\hat{L}^r_{2} - 2n^{-1}\hat{L}^r_{4}
 - 8n^{-2}\hat{L}^r_{0}
 - 8\,n^{-2}\hat{L}^r_{3} + \hat{L}^r_{0} + 3/2\hat{L}^r_{3} + \hat{L}^r_{5} $ \\

45& $   - 2\,n\,\hat{L}^r_{1} - n\,\hat{L}^r_{2} - 3\,n\,\hat{L}^r_{4} + 2\,n^{-1}\,\hat{L}^r_{1} + n^{-1}\,
\hat{L}^r_{2} + 2\,n^{-1}\,\hat{L}^r_{4} - 4\,n^{-2}\,\hat{L}^r_{0} \extraline
- 4\,n^{-2}\,\hat{L}^r_{3} - \hat{L}^r_{0} - 3/2\,\hat{L}^r_{3} - \hat{L}^r_{5} $ \\

47& $   - 1/2\,n\,\hat{L}^r_{5} + 2\,n^{-1}\,\hat{L}^r_{5} $ \\

48& $   - n\,\hat{L}^r_{4} + n^{-1}\,\hat{L}^r_{4} - n^{-2}\,\hat{L}^r_{5} - 1/2\,\hat{L}^r_{5} $ \\

49& $  1/6\,n\,\hat{L}^r_{0} - 7/12\,n\,\hat{L}^r_{3} - 2/3\,\hat{L}^r_{1} - 5\,\hat{L}^r_{2} $ \\

50& $   - 2\,n\,\hat{L}^r_{1} - 1/2\,n\,\hat{L}^r_{2} - 3/2\,\hat{L}^r_{0} - 5/12\,\hat{L}^r_{3} + 1/4
\,\hat{L}^r_{9} $ \\

51& $  1/4\,\hat{L}^r_{2} $ \\

52& $   - 3/4\,n\,\hat{L}^r_{0} - 25/24\,n\,\hat{L}^r_{3} - 8/3\,\hat{L}^r_{2} $ \\

53& $  \hat{L}^r_{0} - 5/4\,\hat{L}^r_{3} $ \\

54& $   - 1/6\,n\,\hat{L}^r_{0} + 5/4\,n\,\hat{L}^r_{3} - 2/3\,\hat{L}^r_{1} + 11\,\hat{L}^r_{2} $ \\

55& $   - 2/3\,n\,\hat{L}^r_{2} - 2/3\,\hat{L}^r_{0} + 1/3\,\hat{L}^r_{3} $ \\

56& $   - 1/2\,\hat{L}^r_{2} $ \\

57& $  2\,n\,\hat{L}^r_{1} + 3/4\,n\,\hat{L}^r_{2} + 5/4\,\hat{L}^r_{0} + 31/24\,\hat{L}^r_{3} - 1/4\,
\hat{L}^r_{9} $ \\

58& $  1/12\,n\,\hat{L}^r_{0} + 1/24\,n\,\hat{L}^r_{3} + 2/3\,\hat{L}^r_{1} + 7/3\,\hat{L}^r_{2} $ \\

59& $  \hat{L}^r_{0} - 3/4\,\hat{L}^r_{3} $ \\

60& $   - 1/12\,n\,\hat{L}^r_{0} - 1/24\,n\,\hat{L}^r_{3} + 2/3\,\hat{L}^r_{1} - 17/3\,\hat{L}^r_{2} $ \\

61& $  \hat{L}^r_{0} + 5/4\,\hat{L}^r_{3} $ \\

62& $   - 1/3\,n\,\hat{L}^r_{2} - 13/3\,\hat{L}^r_{0} - 5/6\,\hat{L}^r_{3} $ \\

63& $   - 2\,\hat{L}^r_{2} $ \\

64& $   - 5/6\,n\,\hat{L}^r_{0} - 7/12\,n\,\hat{L}^r_{3} + 1/8\,n\,\hat{L}^r_{9} - 1/3\,\hat{L}^r_{1}
 - 23/6\,\hat{L}^r_{2} $ \\

65& $   - 4\,n\,\hat{L}^r_{1} - 5/3\,n\,\hat{L}^r_{2} - 14/3\,\hat{L}^r_{0} - 8/3\,\hat{L}^r_{3} + \hat{L}^r_{9}
 $ \\

66& $   - 2/3\,n\,\hat{L}^r_{0} - 1/6\,n\,\hat{L}^r_{5} - 1/12\,n\,\hat{L}^r_{9} - 2/3\,\hat{L}^r_{1}
 - 5\,\hat{L}^r_{2} - 2/3\,\hat{L}^r_{4} $ \\

67& $   - 1/3\,n\hat{L}^r_{0} - 7/6 n\hat{L}^r_{3} - 1/6\,n\hat{L}^r_{5} + 1/6\,n\hat{L}^r_{9}
 - 4/3\hat{L}^r_{1} - 38/3\hat{L}^r_{2} - 2/3\hat{L}^r_{4} $ \\

68& $  1/3\,n\,\hat{L}^r_{0} - 1/6\,n\,\hat{L}^r_{3} + 1/6\,n\,\hat{L}^r_{5} + 1/12\,n\,\hat{L}^r_{9} - 
4/3\,\hat{L}^r_{1} + 6\,\hat{L}^r_{2} + 2/3\,\hat{L}^r_{4} $ \\

69& $   - 2\,\hat{L}^r_{0} + 2\,\hat{L}^r_{3} $ \\

70& $  2/3\,n\,\hat{L}^r_{2} + 8/3\,\hat{L}^r_{0} + \hat{L}^r_{3} + 1/2\,\hat{L}^r_{9} $ \\

71& $   - 1/6\,n\,\hat{L}^r_{3} + 1/12\,n\,\hat{L}^r_{5} + 1/24\,n\,\hat{L}^r_{9} - 1/4\,n\,
\hat{L}^r_{10} - 2/3\,\hat{L}^r_{1} + 1/3\,\hat{L}^r_{4} $ \\

72& $   - 1/6\,n\,\hat{L}^r_{2} - 1/6\,\hat{L}^r_{0} - 1/12\,\hat{L}^r_{3} - 1/4\,\hat{L}^r_{9} - 1/4
\,\hat{L}^r_{10} $ \\

73& $   - 1/6\,n\,\hat{L}^r_{0} + 1/12\,n\,\hat{L}^r_{3} - 1/12\,n\,\hat{L}^r_{5} - 7/24\,n\,
\hat{L}^r_{9} + 2/3\,\hat{L}^r_{1} - 1/3\,\hat{L}^r_{4} $ \\

74& $  1/6\,\hat{L}^r_{3} - 1/2\,\hat{L}^r_{9} - 1/2\,\hat{L}^r_{10} $ \\

75& $  2/3\,n\,\hat{L}^r_{0} + 1/6\,n\,\hat{L}^r_{5} - 5/12\,n\,\hat{L}^r_{9} + 2/3\,\hat{L}^r_{2} + 2/
3\,\hat{L}^r_{4} $ \\

76& $   - 1/3\,n\,\hat{L}^r_{0} + 1/2\,n\,\hat{L}^r_{3} - 1/3\,n\,\hat{L}^r_{5} - 1/3\,n\,\hat{L}^r_{9}
 + 2\,\hat{L}^r_{1} - 1/3\,\hat{L}^r_{2} - 4/3\,\hat{L}^r_{4} $ \\

77& $  2/3\,n\,\hat{L}^r_{2} + 2/3\,\hat{L}^r_{0} + 1/3\,\hat{L}^r_{3} $ \\

78& $  1/6\,n\,\hat{L}^r_{0} - 1/12\,n\,\hat{L}^r_{3} + 1/12\,n\,\hat{L}^r_{5} + 3/8\,n\,\hat{L}^r_{9}
 - \hat{L}^r_{1} - 1/6\,\hat{L}^r_{2} + 1/3\,\hat{L}^r_{4} $ \\

79& $   - 1/3\,\hat{L}^r_{3} $ \\

80& $   - 1/3\,\hat{L}^r_{3} $ \\

81& $   - 1/4\,n\,\hat{L}^r_{9} - 1/4\,n\,\hat{L}^r_{10} $ \\

82& $   - 1/4\,\hat{L}^r_{9} - 1/4\,\hat{L}^r_{10} $ \\

83& $   - 5/6\,n\,\hat{L}^r_{0} - 5/12\,n\,\hat{L}^r_{3} - 1/4\,n\,\hat{L}^r_{5} + 1/8\,n\,\hat{L}^r_{9}
 + 10/3\,n^{-1}\,\hat{L}^r_{0}\extraline
 + 10/3\,n^{-1}\,\hat{L}^r_{3} + 1/3\,\hat{L}^r_{1} - 11/6\,\hat{L}^r_{2} $ \\

84& $   - 4\,n\,\hat{L}^r_{1} - 4/3\,n\,\hat{L}^r_{2} - 2\,n\,\hat{L}^r_{4} + 4\,n^{-1}\,\hat{L}^r_{1} + 4/
3\,n^{-1}\,\hat{L}^r_{2} - 20/3\,n^{-2}\,\hat{L}^r_{0}\extraline
 - 20/3\,n^{-2}\,\hat{L}^r_{3} - 7/3\,\hat{L}^r_{0} - 13/6\,\hat{L}^r_{3} - \hat{L}^r_{5}
 + 1/2\,\hat{L}^r_{9} $ \\

85& $   - 2/3\,n\,\hat{L}^r_{0} - 4/3\,n\,\hat{L}^r_{3} - 1/2\,n\,\hat{L}^r_{5} + 1/4\,n\,\hat{L}^r_{9}
 + 20/3\,n^{-1}\,\hat{L}^r_{0}\extraline
 + 20/3\,n^{-1}\,\hat{L}^r_{3} + 2/3\,\hat{L}^r_{1} - 11/3\,\hat{L}^r_{2} $ \\

86& $   - 1/3\,n\,\hat{L}^r_{0} - 1/2\,n\,\hat{L}^r_{3} - 1/12\,n\,\hat{L}^r_{9} + 2/3\,\hat{L}^r_{1}
 + \hat{L}^r_{2} $ \\

87& $  2\,\hat{L}^r_{0} + 1/3\,\hat{L}^r_{3} - 1/2\,\hat{L}^r_{9} $ \\

88& $   - 4\,n\,\hat{L}^r_{1} - 2/3\,n\,\hat{L}^r_{2} + 4/3\,\hat{L}^r_{0} - 2\,\hat{L}^r_{3} - \hat{L}^r_{9} $ \\

89& $   - 2/3\,n\,\hat{L}^r_{3} - 1/6\,n\,\hat{L}^r_{9} - 2/3\,\hat{L}^r_{1} - \hat{L}^r_{2} $ \\

90& $   - 5/6\,n\,\hat{L}^r_{3} + 1/12\,n\,\hat{L}^r_{5} - 1/8\,n\,\hat{L}^r_{9} + 1/4\,n\,
\hat{L}^r_{10} - 8/3\,\hat{L}^r_{1} - 17/3\,\hat{L}^r_{2}\extraline + 1/3\,\hat{L}^r_{4} $ \\

91& $   - n\,\hat{L}^r_{1} - 1/3\,n\,\hat{L}^r_{2} - 5/6\,\hat{L}^r_{0} - 3/4\,\hat{L}^r_{3} + 1/8\,
\hat{L}^r_{9} + 1/4\,\hat{L}^r_{10} $ \\

92& $   - 1/6\,n\,\hat{L}^r_{0} + 1/12\,n\,\hat{L}^r_{3} - 1/12\,n\,\hat{L}^r_{5} + 1/8\,n\,
\hat{L}^r_{9} + 4/3\,\hat{L}^r_{1} + 7/3\,\hat{L}^r_{2}\extraline
 - 1/3\,\hat{L}^r_{4} $ \\

93& $  1/6\,\hat{L}^r_{3} + 1/4\,\hat{L}^r_{9} + 1/2\,\hat{L}^r_{10} $ \\

94& $   - 1/3\,n\,\hat{L}^r_{0} + 1/6\,n\,\hat{L}^r_{3} - 1/6\,n\,\hat{L}^r_{5} + 1/12\,n\,\hat{L}^r_{9}
 + 10/3\,\hat{L}^r_{1} + 9\,\hat{L}^r_{2} - 2/3\,\hat{L}^r_{4} $ \\

95& $   - 4/3\,n\,\hat{L}^r_{3} - 1/3\,n\,\hat{L}^r_{9} - 8/3\,\hat{L}^r_{1} - 20/3\,\hat{L}^r_{2} $ \\

96& $  2\,\hat{L}^r_{0} + 1/3\,\hat{L}^r_{3} - 1/2\,\hat{L}^r_{9} $ \\

97& $   - 2/3\,n\,\hat{L}^r_{3} + 1/12\,n\,\hat{L}^r_{5} - 1/8\,n\,\hat{L}^r_{9} - 5/3\,\hat{L}^r_{1}
 - 9/2\,\hat{L}^r_{2} + 1/3\,\hat{L}^r_{4} $ \\

98& $  \hat{L}^r_{0} + 1/6\,\hat{L}^r_{3} - 1/4\,\hat{L}^r_{9} $ \\

99& $   - 4\,n\,\hat{L}^r_{1} - 2/3\,n\,\hat{L}^r_{2} - 8/3\,\hat{L}^r_{0} - 8/3\,\hat{L}^r_{3} $ \\

100& $   - 1/6\,n\,\hat{L}^r_{0} + 1/12\,n\,\hat{L}^r_{3} - 1/12\,n\,\hat{L}^r_{5} - 3/8\,n\,
\hat{L}^r_{9} + 1/3\,\hat{L}^r_{1} - 1/6\,\hat{L}^r_{2}\extraline
 - 1/3\,\hat{L}^r_{4} $ \\

101& $  1/6\,n\,\hat{L}^r_{0} - 1/12\,n\,\hat{L}^r_{3} + 1/12\,n\,\hat{L}^r_{5} - 1/8\,n\,\hat{L}^r_{9}
 - 1/3\,\hat{L}^r_{1} + 1/6\,\hat{L}^r_{2} + 1/3\,\hat{L}^r_{4} $ \\

102& $   - 1/3\,n\,\hat{L}^r_{0} - 2/3\,n\,\hat{L}^r_{3} - 1/4\,n\,\hat{L}^r_{5} + 1/4\,n\,
\hat{L}^r_{10} + 10/3\,n^{-1}\,\hat{L}^r_{0} \extraline
+ 10/3\,n^{-1}\,\hat{L}^r_{3} - 2/3\,\hat{L}^r_{1} - 4/3\,\hat{L}^r_{2} $ \\

103& $   - n\,\hat{L}^r_{1} - 1/3\,n\,\hat{L}^r_{2} - 1/2\,n\,\hat{L}^r_{4} + n^{-1}\,\hat{L}^r_{1} + 1/
3\,n^{-1}\,\hat{L}^r_{2} - 5/3\,n^{-2}\,\hat{L}^r_{0}\extraline
 - 5/3\,n^{-2}\,\hat{L}^r_{3} - 1/3\,\hat{L}^r_{0} - 2/3\,\hat{L}^r_{3} - 1/4\,
\hat{L}^r_{5} + 1/4\,\hat{L}^r_{10} $ \\

104& $  1/12\,n\,\hat{L}^r_{0} - 1/24\,n\,\hat{L}^r_{3} + 1/24\,n\,\hat{L}^r_{5} + 5/48\,n\,
\hat{L}^r_{9} - 1/6\,\hat{L}^r_{1} + 1/12\,\hat{L}^r_{2}\extraline + 1/6\,\hat{L}^r_{4} $ \\

105& $  1/6\,n\,\hat{L}^r_{0} + 13/12\,n\,\hat{L}^r_{3} + 1/12\,n\,\hat{L}^r_{5} + 1/6\,n\,\hat{L}^r_{9}
 - 8/3\,n^{-1}\,\hat{L}^r_{0} - 8/3\,n^{-1}\,\hat{L}^r_{3}\extraline
 + 3\,\hat{L}^r_{1} + 11/2\,\hat{L}^r_{2} - 5/3\,\hat{L}^r_{4} $ \\

106& $  6\,n\,\hat{L}^r_{1} + 5/3\,n\,\hat{L}^r_{2} + 2\,n\,\hat{L}^r_{4} - 4\,n^{-1}\,\hat{L}^r_{1} - 2/3
\,n^{-1}\,\hat{L}^r_{2} + 16/3\,n^{-2}\,\hat{L}^r_{0} \extraline
+ 16/3\,n^{-2}\,\hat{L}^r_{3} + 5/3\,\hat{L}^r_{0} + 25/6\,\hat{L}^r_{3} + 3/4\,
\hat{L}^r_{9} $ \\

107& $  1/3\,n\,\hat{L}^r_{0} + 7/6\,n\,\hat{L}^r_{3} + 1/2\,n\,\hat{L}^r_{5} + 1/4\,n\,\hat{L}^r_{9} - 
16/3\,n^{-1}\,\hat{L}^r_{0}\extraline
 - 16/3\,n^{-1}\,\hat{L}^r_{3} + 2/3\,\hat{L}^r_{1} + 7/3\,\hat{L}^r_{2} $ \\

108& $  4\,n\,\hat{L}^r_{1} + 2/3\,n\,\hat{L}^r_{2} + 2\,n\,\hat{L}^r_{4} - 4\,n^{-1}\,\hat{L}^r_{1} - 2/3
\,n^{-1}\,\hat{L}^r_{2} + 16/3\,n^{-2}\,\hat{L}^r_{0}\extraline
 + 16/3\,n^{-2}\,\hat{L}^r_{3} + 2/3\,\hat{L}^r_{0} + 7/3\,\hat{L}^r_{3} + \hat{L}^r_{5}
 + 1/2\,\hat{L}^r_{9} $ \\

109& $   - 1/6\,n\,\hat{L}^r_{9} $ \\

110& $  1/3\,n\,\hat{L}^r_{0} - 1/6\,n\,\hat{L}^r_{3} + 1/6\,n\,\hat{L}^r_{5} + 1/12\,n\,\hat{L}^r_{9}
 - 2/3\,\hat{L}^r_{1} + 1/3\,\hat{L}^r_{2} + 2/3\,\hat{L}^r_{4} $ \\

111& $   - 1/3\,n\,\hat{L}^r_{0} + 1/6\,n\,\hat{L}^r_{3} - 7/12\,n\,\hat{L}^r_{9} + 2/3\,\hat{L}^r_{1}
 - 1/3\,\hat{L}^r_{2} $ \\

112& $  1/6\,n\,\hat{L}^r_{5} + 2/3\,\hat{L}^r_{4} $ \\

114& $   - 4/3\,n\,\hat{L}^r_{9} $ \\

115& $  1/3\,n\,\hat{L}^r_{9} $ \\
\end{longtable}

\end{document}